\documentclass[letterpaper,twocolumn,10pt]{article}
\usepackage{usenix2019_v3}

\usepackage{tikz}
\usepackage{amsmath}

\usepackage{filecontents}

\usepackage{subfiles}
\usepackage{url}
\usepackage{hyperref}
\usepackage{xurl}
\usepackage{graphicx}
\usepackage{subcaption}
\usepackage{booktabs}
\usepackage{appendix}

\usepackage{color}

\begin{document}

\date{}

\title{\Large \bf DoLLM: How Large Language Models Understanding Network Flow Data\\
to Detect Carpet Bombing DDoS\\
}

\author{
{\rm Qingyang Li}\textsuperscript{\textnormal\textdagger}, 
{\rm Yihang Zhang}\textsuperscript{\textnormal\textdagger},
{\rm Zhidong Jia}\textsuperscript{\textnormal\textdagger}, 
{\rm Yannan Hu}\textsuperscript{\textnormal\textasteriskcentered}, 
{\rm Lei Zhang}\textsuperscript{\textnormal\textasteriskcentered}, \\
{\rm Jianrong Zhang}\textsuperscript{\textnormal\textdaggerdbl}, 
{\rm Yongming Xu}\textsuperscript{\textnormal\textdaggerdbl},
{\rm Yong Cui}\textsuperscript{\textnormal\textbullet}, 
{\rm Zongming Guo}\textsuperscript{\textnormal\textdagger},
{\rm Xinggong Zhang}\textsuperscript{\textnormal\textdagger}
\\
\textsuperscript{\textnormal\textdagger}Peking University, 
\textsuperscript{\textnormal\textbullet}Tsinghua University,
\textsuperscript{\textnormal\textasteriskcentered}Zhongguancun Laboratory, 
\textsuperscript{\textnormal\textdaggerdbl}Unicom Digital Tech
}


\maketitle


\begin{abstract}
It is an interesting question Can and How Large Language Models (LLMs) understand non-language network data, and help us detect unknown malicious flows.   
This paper takes Carpet Bombing as a case study and shows how to exploit LLMs' powerful capability in the networking area.

Carpet Bombing is a new DDoS attack that has dramatically increased in recent years, significantly threatening network infrastructures. It targets multiple victim IPs within subnets, causing congestion on access links and disrupting network services for a vast number of users. Characterized by low-rates, multi-vectors, these attacks challenge traditional DDoS defenses.

We propose DoLLM, a DDoS detection model utilizes open-source LLMs as backbone.
By reorganizing non-contextual network flows into Flow-Sequences and projecting them into LLMs semantic space as token embeddings, DoLLM leverages LLMs' contextual understanding to extract flow representations in overall network context.
The representations are used to improve the DDoS detection performance.

We evaluate DoLLM with public datasets \textit{CIC-DDoS2019} and real \textit{NetFlow} trace from Top-3 countrywide ISP. The tests have proven that DoLLM possesses strong detection capabilities. Its F1 score increased by up to 33.3\% in zero-shot scenarios and by at least 20.6\% in real ISP traces.


\end{abstract}
\section{Introduction}
Carpet Bombing is a new type of DDoS attack that has rapidly increased over recent years~\cite{carpet_a10,carpet_corero,carpet_netscout,carpet_radware,carpet_huawei,carpet_FastNetMon}.
According to the Reports~\cite{carpet_South_african}, In 2019 the largest ISP in South Africa was attacked by Carpet Bombing, severing internet connections for all its users for entire day. Countries like Brazil, the United States, and China frequently suffer from Carpet Bombing attacks, and more than 400,000 carpet-bombing attacks since July 2023~\cite{carpet_netscout}. Unlike traditional DDoS, Carpet Bombing seemly attacks multiple victim IPs within IDCs, public clouds, CDNs, and even ISPs. But its real intend behind is to {\it flood their access links}~\cite{carpet_DRDoS}. This pose significant challenges to current DDoS detection systems.    


\begin{figure}
    \centering
    \includegraphics[width=0.95\linewidth]{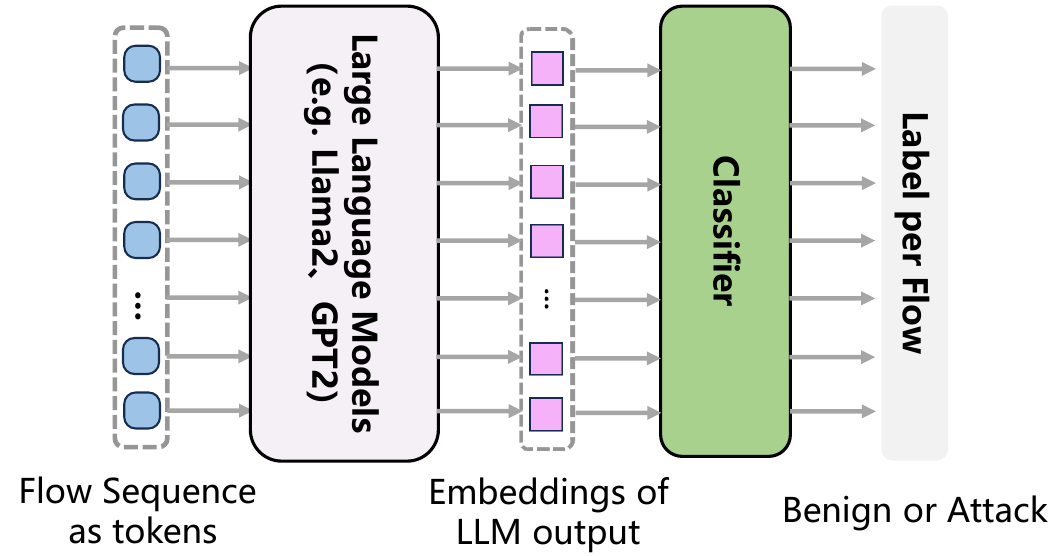}
    \caption{Diagram of DoLLM  to detect Carpet Bombing by representation learning. It aligns network flows data to LLM semantic space for better exploiting flow correlations.}
    \label{intro}

    \vspace{-0.5cm}
\end{figure}

\textbf{It's reluctant to detect Carpet Bombing efficiently relying solely on flow's explicit features}, such as traffic, IPs, protocols, etc.
Firstly, 73.19\% of Carpet Bombing~\footnote{Conversely, some Carpet Bombing attacks still trigger host-side detection thresholds, but this type of attack has obvious traffic features, and is out of scope of this paper.} attacks are low-rate~\cite{carpet_huawei}, meaning each victim IP receives only a small amount of attack traffic that fails to trigger the host-target traffic thresholds.
Secondly, 86.96\% of Carpet Bombing attacks use multiple vectors~\cite{carpet_huawei}, causing the explicit features of attack flows to blend with those of benign flows. The methods typically rely on setting rules~\cite{liu2021jaqen} or feature-engineering machine learning~\cite{accturbo-alcoz2022aggregate,net-beacon-zhou2023efficient,wichtlhuber2022ixp} fail to accurately distinguish Carpet Bombing from benign traffic. Jaqen~\cite{liu2021jaqen} and ACC-turbo~\cite{accturbo-alcoz2022aggregate} collaterally drop benign traffic with 73.19\% and 19.98\% respectively in Carpet Bombing scenarios~\cite{accturbo-alcoz2022aggregate}.
There are also some link-flood detection methods~\cite{link_kang2013crossfire,link_xing2021ripple}. But they require rerouting traffic for detection~\cite{link_xing2021ripple,link_lee2013codef,link_liaskos2016novel}, which is often impractical to mitigate Carpet Bombing attack on access links.

Inspired by the success of large language models (LLMs) in mining textual context information, it is desirable to detect Carpet Bombing by exploiting LLMs' powerful capabilities of {\it semantic representation, adaptability and transfer learning}.
It performs \textbf{representation learning} on the flows and explore their latent correlations features. This idea aims to move away from traditional hand-crafted feature engineering to enhance the accuracy of Carpet Bombing detection.

However, using LLMs for representation learning on non-language network flow data are totally new challenges:
\begin{enumerate}
    \item \textbf{Network flow data is a modality different from natural language}, which LLMs cannot directly process. It is necessary to align the modalities of network flow data to LLMs semantic space.
    \item Network flows arrive randomly from numerous users and devices. \textbf{There exists not flow context as language}. This limits LLMs' ability to mine inter-flow correlations.
    \item The Decoder-Only architecture of LLMs is \textbf{well-suited for generation tasks but may lead to information loss in representation learning}. Therefore, the structure of LLMs needs to be adjusted to better accommodate representation learning.
\end{enumerate}

Therefore, we propose DoLLM (D\textbf{Do}S Detection based on \textbf{L}arge \textbf{L}anguage \textbf{M}odels), a Carpet Bombing detection model that \textbf{utilizes open-source LLMs as its backbone}. The pipeline is illustrated in Figure~\ref{intro}.
We consider flow-level DDoS detection as a Token\footnote{A token is a basic unit in text, which can be considered a word or a character.} Classification problem. The non-contextual network flows are organized into contextual Flow Sequence, treating each flow as the smallest unit in text.
The network flow data is then translated into tokens by the proposed Flow Tokenizer. 
These tokens are processed to extract representations by the LLMs' backbone, producing high-dimensional embeddings in the LLMs' semantic space. Finally, these embeddings are used for binary classification to identify attack or benign flows. 

We evaluated the performance of DoLLM with public CIC-DDoS2019~\cite{cic_sharafaldin2019developing} and a real NetFlow trace from Top-3 countrywide ISP. DoLLM demonstrated the best detection performance across all scenarios. Notably, in zero-shot scenarios, DoLLM's F1 score reached 0.964, an {\it improvement upmost of 33.3\%} over other methods, indicating DoLLM's strong capability to detect new types of attacks. In real ISP Trace, DoLLM's F1 score was {\it at least 20.6\% higher} than other methods, showcasing its significant advantage in real-world scenarios. We also conducted ablation and parameter experiments to validate the functionality of DoLLM's core modules.


In summary, the contributions of this paper are as follows:

\begin{itemize}
    \item To our knowledge, this is \textbf{the first effort to explore LLMs representation learning capabilities in understanding DDoS network flow data}.
    \item We introduced the Flow Sequentializer, which organizes temporally unordered flows into a \textbf{structured Flow Sequence}, facilitating the mining of correlations between flows. 
    \item We conceptualize flow information as a new modality and designed multimodal interfaces, Flow Tokenizer and Classification Projection, to \textbf{align the feature space of flows with that of natural language}, enabling LLMs to perform flow detection tasks.
    \item We modeled the Flow-level DDoS detection task as a \textbf{Token Classification task}, allowing the LLMs to extract representations for each flow while exploring the inter-flow correlations.
\end{itemize}


\section{Background and Motivation}
\label{bg_moti}

\begin{figure*}[htbp]
    \centering
        \begin{subfigure}[b]{0.31\linewidth}
            \includegraphics[width=0.9\textwidth]{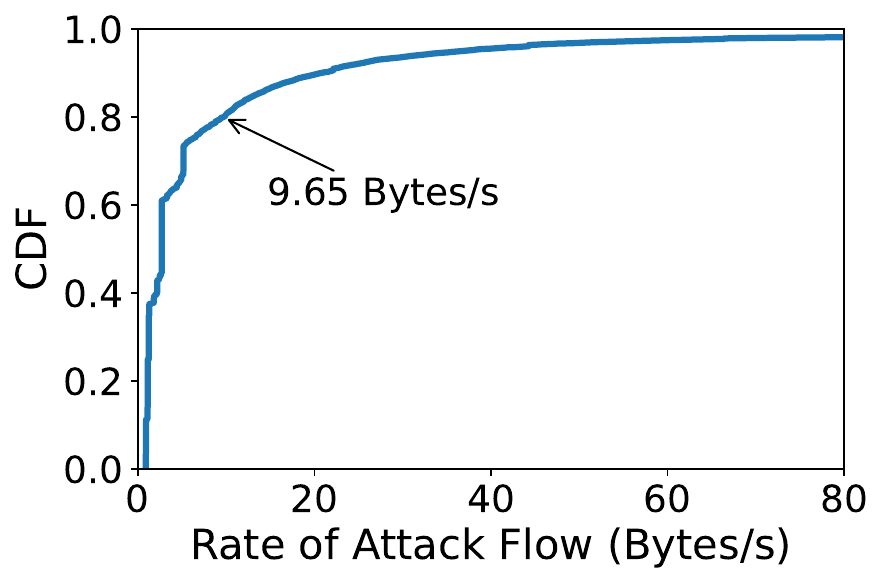}
            \caption{CDF of attack flow byte-rate.}
            \label{moti-carp-rate}
        \end{subfigure}
        \hfill
        \begin{subfigure}[b]{0.31\linewidth}
            \includegraphics[width=0.9\textwidth]{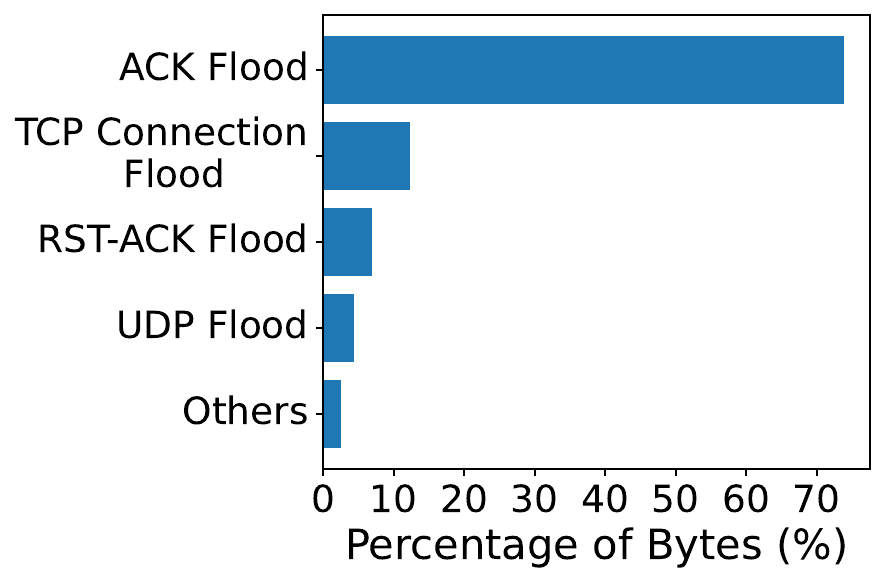}
            \caption{Distribution of attack vectors.}
            \label{moti-attack-type}
        \end{subfigure}
        \hfill
        \hspace{2mm}
        \begin{subfigure}[b]{0.31\linewidth}
            \includegraphics[width=0.9\textwidth]{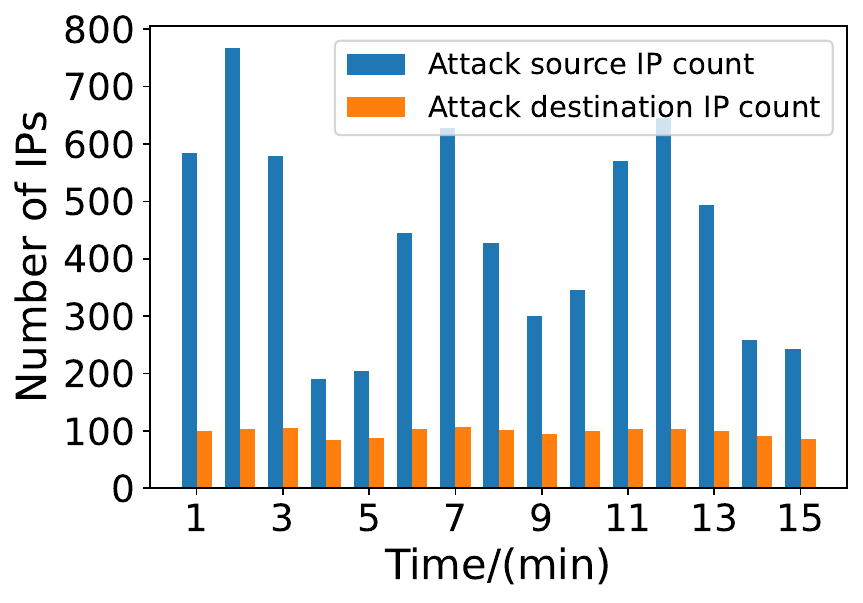}
            \caption{Attack Src/Dest IP counts.}
            \label{moti-attack-num}
        \end{subfigure}
        \caption{Carpet Bombing features with lower-rate single flow traffic, multiple attack vectors and many-to-many attack natures.}
        \vspace{-3mm}
        \label{moti-carpet_feature}
\end{figure*}

\subsection{Large Language Model}


The release of models such as GPT4~\cite{achiam2023gpt}, Llama2~\cite{touvron2023llama}, and GLM~\cite{zeng2022glm} led to rapid advancements in large language models (LLMs). LLMs are built on the Transformer~\cite{transformer} architecture, featuring over a billion neural network parameters, and have been pretrained on massive text corpora using autoregressive methods. This design allows LLMs to  generate text by processing input sequences into {\it high-dimensional token embeddings}, further enhanced with positional encoding to capture the sequential and contextual correlations between tokens through multiple layers of Transformer blocks. As a result, LLMs demonstrate profound capabilities in handling a variety of natural language processing (NLP) tasks, including dialogue, summarization, and translation, markedly improving both professional workflows and daily life interactions.

LLMs have been used as the backbone to solve specific tasks in many different areas besides NLP for its strong generalized capability, where the greatest challenge lies in enabling LLMs to \textbf{comprehend non-natural language data}, such as images, and videos.  
Multimodal Large Language Models (MLLMs)~\cite{survey_zhang2024mm} aim to combine information from multiple modalities to improve their understanding and performance across various tasks. 
They have already been applied in fields such as imaging~\cite{image_li2023blip,image_liu2023llava,autoregressive_awadalla2023openflamingo}, video~\cite{video_li2023videochat,video_maaz2023video,video_li2023llama}, time series~\cite{time_jin2023time,time_liu2024autotimes,time_zhou2024one}, and networks~\cite{network_wu2024large}. In MLLMs, the \textit{Input Projector} and \textit{Output Projector} are critical modules that align features from other modalities to the textual feature space, enabling LLMs to process~\cite{survey_zhang2024mm}. However, no previous works have tried LLMs to understand raw network flow data that have not been exposed to LLMs during pretraining.

To harness the potential of LLMs in networking, there are primarily two categories of approaches: {\it generative models} and {\it representation learning}. Generative models generates token sequences autoregressively~\cite{autoregressive_aiello2023jointly,autoregressive_awadalla2023openflamingo,time_liu2024autotimes} and converting them into content of corresponding modalities. Others utilize representation learning~\cite{time_jin2023time,time_zhou2024one}, treating LLMs as feature extractors for token sequences and using the captured representations for specific downstream tasks. We argue that the \textbf{generative models is not suitable for certain non-generative tasks}, such as classification, and clustering. Firstly, these tasks are solely concerned with the features of the input sequence itself and do not require the generation of additional content. Secondly, the generative models requires multiple rounds of inference through LLMs, which is inefficient~\cite{network_wu2024large}, whereas representation learning only need a single round of inference to extract features. Therefore, in this paper, we exploits LLMs in a \textit{representation learning} way to detect Carpet Bombing.

\subsection{Characteristics and Challenges of Carpet Bombing}

Carpet Bombing~\cite{carpet_DRDoS} has become one of the most severe \textbf{threats to network infrastructure}, congesting the access links by flooding. It is a kind of many-to-many attacks that is hard to detect, launching from numerous compromised devices, and targeting a large number of victim IPs across one or more subnets. Carpet Bombing frequently targets large-scale network infrastructures such as IDCs, public clouds, CDNs, and even ISPs, affecting the normal communication of massive numbers of users. Moreover, the intensity and scale of Carpet Bombing attacks are continually increasing. The aggregate attack traffic can exceed 100 Gbps and may cover 200 to 600 Class C network segments~\cite{carpet_huawei}.

Carpet Bombing is characterized by the nature of \textbf{low-rate, multi-attack-vectors and many-sources-to-many-destinations} nature. The conclusions were drawn from network traffic captured during the period of real Carpet Bombing. This event occurred in August 2022, targeting hundreds of users from multiple Class C subnets, and lasted for over seven hours. We obtained attack packets from a 26-minute period and created a dataset for analysis. The attack packets were captured by a DDoS scrubbing device, and we extracted the flow information based on the five-tuple of \textit{source IP}, \textit{destination IP}, \textit{source port}, \textit{destination port}, and \textit{protocol}. We analyzed the byte-rate, the distribution of attack vectors, and the characteristics of their IP counts, as shown in Figure ~\ref{moti-carpet_feature}.

As shown in Fig.~\ref{moti-carp-rate}, Carpet Bombing features \textbf{low rate} for every attack flow, with 80\% of the flows having traffic less than 9.65 Bytes/s. As the aggregate attack traffic is shared among all the victims, each victim receives a relatively small amount of attack traffic. In 73.19\% of carpet bombing attacks, the attack data rate on each victim IP is so low that it fails to trigger host-side detection thresholds~\cite{carpet_huawei}. The attackers further use low-rate flows to achieve the desired effect, making the attacks more covert.


As shown in Fig.~\ref{moti-attack-type}, Carpet Bombing also features \textbf{various attack vectors} during one attack event, consisting of more than five types of attack vectors, primarily initiated by transport layer protocols. The same observation is reported in ~\cite{carpet_huawei}, with 86.96\% of Carpet Bombing attacks composed of a mixture of different types of DDoS attacks, since Carpet Bombing can utilize any protocol and application to conduct.

As shown in Fig.~\ref{moti-attack-num}, Carpet Bombing, as its definition, exhibits \textbf{many-to-many} natures. We divided the dataset into one-minute intervals and calculated the number of source and destination IPs in each interval for the last 15 minutes of the dataset. The results show that the attack is launched from over 200 different source compromised devices and targeted at over 100 different destination IP addresses in multiple Class C subnets, indicating a wide coverage of Carpet Bombing.


For the aforementione reasons, Carpet Bombing challenges current DDoS detection methods. Most host-side detection methods fail because they are not triggered, making detection possible only at the access link. Existing Link Flood detection methods, like those in~\cite{link_xing2021ripple,link_lee2013codef,link_liaskos2016novel}, mostly require rerouting, hard to implement at the access link. Methods based on programmable switches~\cite{liu2021jaqen, accturbo-alcoz2022aggregate, net-beacon-zhou2023efficient} and core network detection centers~\cite{wichtlhuber2022ixp} mostly rely on hand-craft feature engineering, struggling against multi-vectors attack that resembles benign flow. In experiments, ACC-Turbo~\cite{accturbo-alcoz2022aggregate} and Jaqen~\cite{liu2021jaqen} had high benign packet drop rates of 19.98\% and 73.19\%~\cite{accturbo-alcoz2022aggregate}, fails to accurately identify benign and malicious flows.

\subsection{Observations of Flows-Correlations}

Since Carpet Bombing is a kind of many-to-many attack, focusing \textbf{solely on the features of individual flows} makes it difficult to accurately distinguish between benign and attack flows, as their explicit features can be similar. Therefore, we come up with the question that \textit{is it possible to exploit multi-flows correlation for detection?} In DDoS detection, it is essential to consider the correlations between flows, which can be characterized by their similarity. Flows generated by the same application or attack script often exhibit similar characteristics. Leveraging this correlations can enhance the accuracy of DDoS detection and traffic classification tasks~\cite{corr_yu2011discriminating,corr_zhang2012network}.


To answer this question, we analyze the \textbf{inter-flows correlations} using a simulated Carpet Bombing dataset. We use a MAWI~\cite{mawi_kenjiro2000traffic} dataset collected in a backbone network (January 1, 2024) as the benign background traffic. We extracted Flow information according to the five-tuple, and mixed it with the attack flow mentioned above to simulate a Carpet Bombing. As described in \S~\ref{design_Flow_Tokenizer}, we calculated nine features for each flow and normalized them to form a nine-dimensional feature vector. We then calculated the cosine similarity between attack flows, as well as between benign flows.

\begin{figure}
    \centering
    \includegraphics[width=0.8\linewidth]{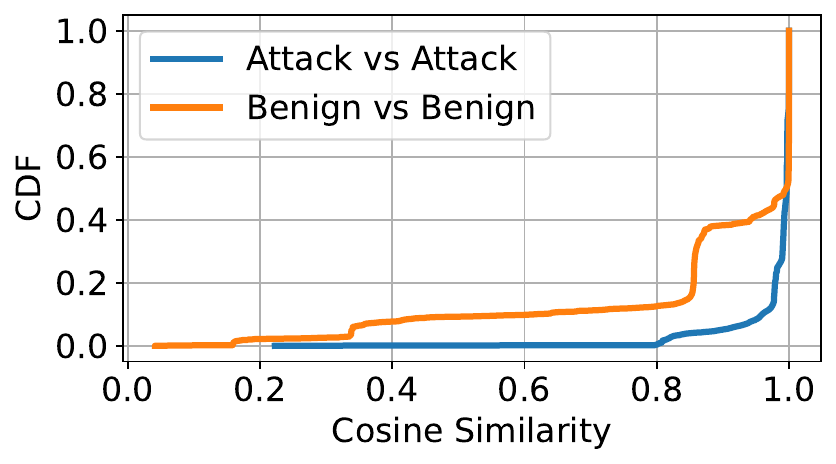}
    \caption{There exists stronger correlations among malicious-flows than among benign-flows.}

    \label{moti-corr}
\end{figure}

As shown in Fig.~\ref{moti-corr}, compared with the cosine similarities between benign flows, \textbf{the cosine similarities between attack flows are generally stronger}, with 80\% of the similarities above 0.95. This is because attack flows within the same attack event are typically generated by the same set of scripts, hence they exhibit similar characteristics. In contrast, benign flows are generated by various normal applications, and thus their features are distinct from each other, explaining for the generally weaker cosine similarities inter benign flows.


The above observation give us a hint, since there exists strong similarity inter attack flows, \textit{if we can exploit the flow-correlations using LLMs, then it is possible to capture the general representation of "malicious flow clusters", and more precisely distinguish the subtle pattern differences between attack flows and benign flows}. This approach enhances the accuracy of detecting Carpet Bombing attacks.

\subsection{Our Insight}

Benefiting from extensive pretraining on large text corpora and self-attention mechanism, \textbf{LLMs are highly effective at discovering correlations between tokens}, which in turn enhances their ability to understand context. This inspires us to view each network flow as a token, organizing different flows into sequences. 
By employing LLMs to process the organized sequences, we can \textbf{capture the correlations} between flows, \textbf{extracting the representations} of each flow for \textbf{further classification} to better detect Carpet Bombing. 

However, leveraging LLMs to explore the correlations between flows for Carpet Bombing detection is challenging:

    \textbf{Challenge 1:} Unlike natural language texts, which have defined grammatical structures, network flows collected are typically ordered by arrival time and \textbf{do not have a temporal dependency on one another}. Therefore, it is necessary to reorganize these flows under specific rules to better discover correlations between them.
    
    \textbf{Challenge 2:} Flows represent \textbf{a modality different from natural language}, which LLMs cannot directly process. It is essential to align these two modalities, mapping flow information into the Semantic space of LLMs, allowing LLMs to handle flow data.
    
    \textbf{Challenge 3:} LLMs were originally \textbf{designed for text generation, not for representation learning}. Therefore, adjusting their structure to better suit representation learning of flow information is required.

\vspace{-0.5cm}

\begin{figure*}
\vspace{-2mm}
    \centering
    \includegraphics[width=0.7\linewidth]{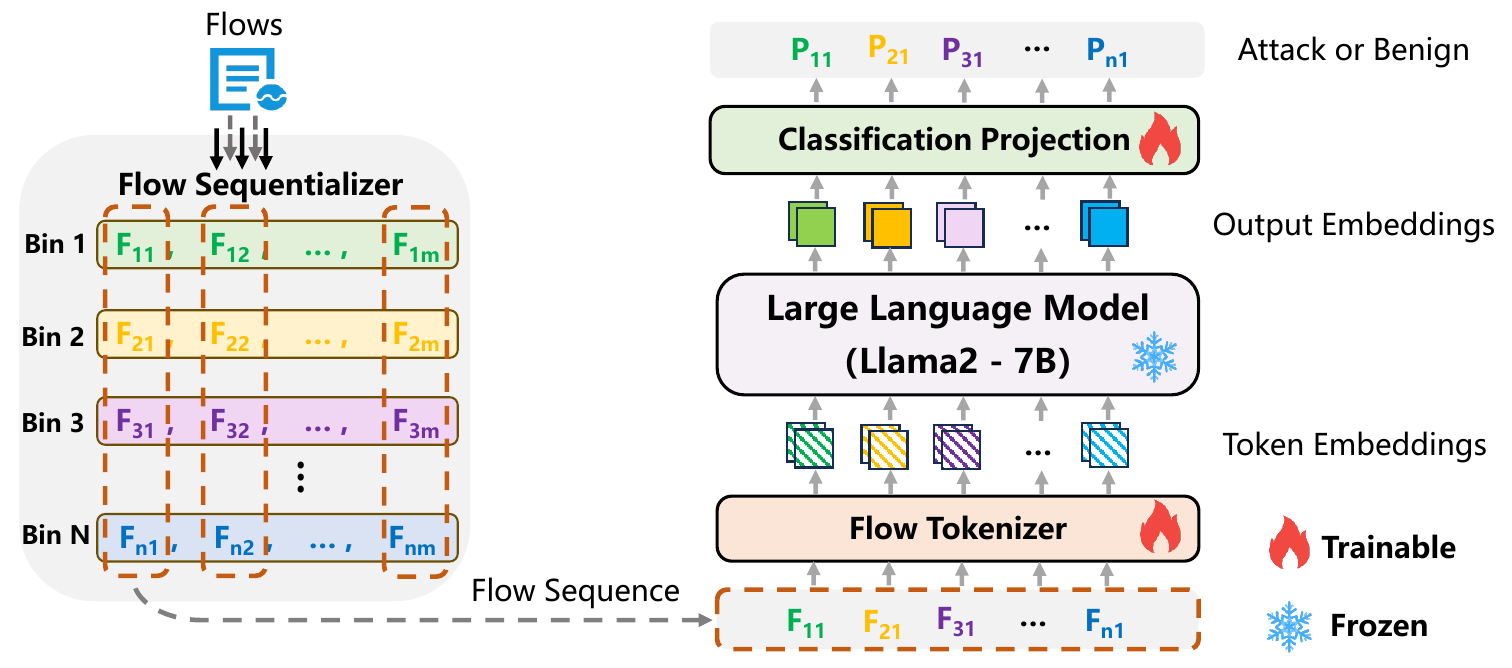}
    \caption{The overall architecture of DoLLM. It consists of four parts: Flow Sequentializer, Flow Tokenizer, Bidirectional Self-Attention LLMs, and Classification Projection.}
    \label{overview_dollm}
    \vspace{-4mm}
\end{figure*}

\section{Overview}
\label{overview}

\textbf{DoLLM is a DDoS detection model that leverages LLMs as backbone.} It identifies correlations among network flows and extracts representations of each flow to detect DDoS attacks. Considering the large number of packets in the access links, extracting packet-level features, particularly payload data, is impractical. Therefore, we utilize features derived from five-tuple statistics to describe flows. As such, DoLLM not only adapts well to the common NetFlow~\cite{netflow_claise2004cisco} detection methods used in current ISPs but can also be applied to flow information gathered directly by devices~\cite{cic_sharafaldin2019developing}. 

The design of DoLLM is inspired by the classic problem of Token Classification in the field of NLP\cite{Ner_li2020survey,pos_schmid1994part} where each token in the text is labeled with a tag. Transformer based Language models output embeddings for each token in the text~\cite{Ner_li2020survey}, and these embeddings have already incorporated the correlations between tokens via the self-attention mechanism.

\textbf{Therefore, we modeled the DDoS detection issue as a Token Classification problem.} Each flow is viewed as a token, and multiple flows are organized into an ordered sequence. Utilizing LLMs, the embeddings for each flow is extracted for benign/attack flows classification. These embeddings are not only represent single-flow's explicit features,  but also considers the correlations between flows.
This method could effectively enhances detection efficiency and performance. 

The overall architecture of DoLLM is shown in Figure ~\ref{overview_dollm}, which consists of the following four modules:

\begin{itemize}
    \item \textbf{Flow Sequentializer (\S~\ref{design_Flow_Sequence_Generator}):} This module generates contextually Flow Sequences from temporally unordered flows using specific rules.
    
    \item \textbf{Flow Tokenizer (\S~\ref{design_Flow_Tokenizer}):} A learnable module that normalizes the features of each flow in the Flow Sequence and encodes them into Token Embeddings that can be processed by large language models.
    
    \item \textbf{Bidirectional Self-Attention LLMs (\S~\ref{design_llm}):} We use Llama2-7B~\cite{touvron2023llama} as the backbone and the network parameters of this module are frozen. The causal mask in LLM is removed, allowing the LLMs to explore the correlations between any two flows in the Flow-Sequence through bidirectional self-attention. This process conducting representation learning to extract high-dimensional flows' features in the LLM semantic space.
    
    \item \textbf{Classification Projection (\S~\ref{design_class}):} A learnable module that projects the LLM's output embeddings of each flow, into a vector space more conducive to classification tasks. Then it performs binary classification to determine whether the flows are benign or attack.
\end{itemize}

\section{DoLLM Design}
\label{design}
In this section, we describe the design details of DoLLM. And the training and inference end-to-end pipeline are presented.

\begin{figure*}
\vspace{-2mm}
    \centering
    \includegraphics[width=0.8\linewidth]{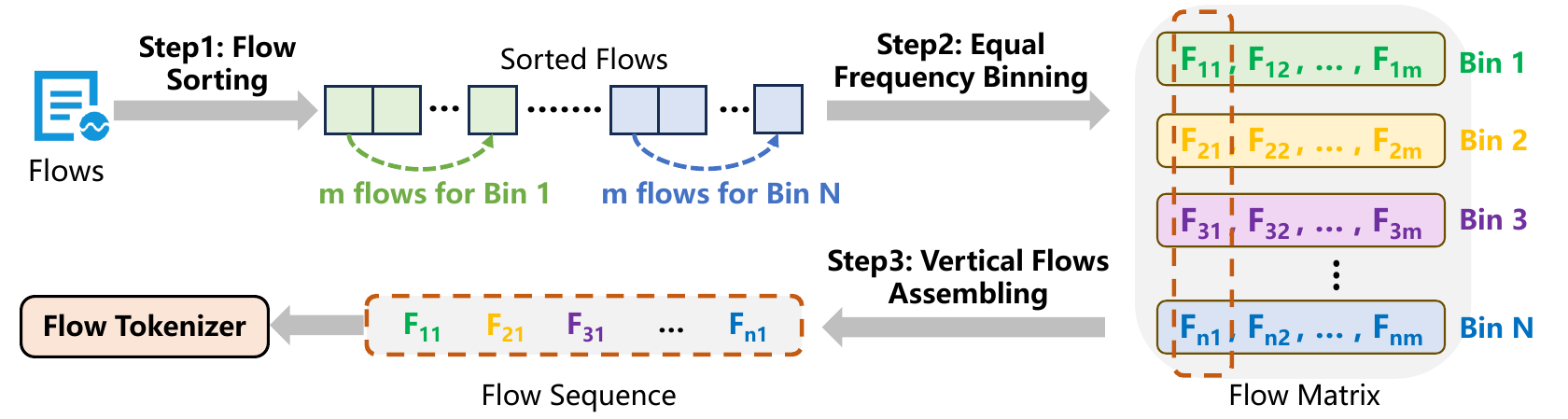}
    \caption{Illustration of the workflow for Flow Sequentializer which generates contextual Flow Sequences from temporally unordered raw flows.}
    \label{Flow_Sequence_Generator}
    \vspace{-0.3cm}
\end{figure*}

\subsection{Flow Sequentializer}
\label{design_Flow_Sequence_Generator}
To enable LLMs to explore the correlations between flows, it is necessary to organize flows into sequences. However, simply dividing sequences by the order of their arrival time is useless.
Since the flows usually originate from different users or services, their arrival times are independent of one another. For attack flows, although they may be launched or stopped synchronously, but due to complex network topology, congested links, and huge amounts of mixed benign traffic, 
attack flows are not necessary to arrive at the same time with clear temporal correlation.

It's a problem for LLMs to extract representations from such non-contextual sequences. Due to the presence of positional encoding, LLMs are highly sensitive to the position of each token in the input sequence.
Therefore, \textbf{we designed the Flow Sequentializer to create contextual Flow Sequences from temporally unordered flows.} The architecture of the Flow Sequentializer is shown in Figure ~\ref{Flow_Sequence_Generator}.

There are two key considerations in the design of the Flow Sequentializer: 

\begin{enumerate}
    \item To ensure coverage of all flows in detection, each flow must appear in a Flow Sequence. Additionally, each flow should ideally appear only once in a Flow Sequence to avoid redundant inference on the same flow.
    \item Flows at the same positions in the Flow Sequence must have similar properties, otherwise, it would be impossible to learn a fixed pattern.
\end{enumerate}

To address these two issues, we aim to develop a set of rules for generating Flow Sequences, drawing inspiration from the grammatical rules of natural language. The generation of Flow Sequence consists of the following three steps:

    \textbf{Step 1: Flow Sorting,} which orders the flows received by DoLLM according to the priority of average packet length, total packets, protocol, source port, and destination port to make them orderly. We selected these five features because they are closely related to the higher-level applications of the flows and can accurately reflect the patterns of the flows.

    \textbf{Step 2: Equal Frequency Binning,} which aims to distribute flows into various bins, ensuring that each bin contains the same number of flows. Since the total number of flows may not divide evenly by the number of bins, we need to predefine the number of flows in each bin. To accommodate this, bins with smaller index will contain one additional flow. Suppose we have \(F\) flows that need to be evenly divided across \(N\) bins. Then, the number of flows \(m_i\) in the bin with index \(i\) can be expressed as follows:
        \begin{equation}
        m_i = 
        \begin{cases} 
        q + 1, & i \leq r \\
        q,     &  i > r
        \end{cases}
        \label{eq:ni}
        \end{equation}
        Where \( q = \left\lfloor \frac{F}{N} \right\rfloor\), \(r = F\mod N\). Each flow fills the bins sequentially in the order in which they were sorted by Step 1, first filling the first bin, then the second bin, and so on. And We duplicate the last flow in the bins with fewer flows to ensure that all bins contain an equal number of flows. This allows the flows in all bins to be neatly organized into a matrix, which we refer to as the Flow Matrix. Through the above binning process, we can group flows with similar explicit characteristics together, treating them as tokens with similar semantic meanings.
        
    \textbf{Step 3: Vertical Flows Assembling,} which involves selecting flows from the same position across different bins to form a Flow Sequence, specifically by selecting columns from the Flow Matrix. The length of each selected Flow Sequence is \(N\), and the number of sequences is \(\left\lceil \frac{F}{N} \right\rceil\). This approach ensures that each flow appears approximately once in the Flow Sequence, preventing redundant detection for each flow.
    
    It's worth noting that for the model's inference process, generating the Flow Sequence as described above and passing it to the LLMs for inference is sufficient. This allows LLMs to perform representation learning for multiple flows in a single inference step, effectively increasing the detection speed. However, for the training process, following this method alone may result in too few Flow Sequence samples, potentially leading to non-convergence of the model. Therefore, in addition to the limited number of Flow Sequence samples generated as described, we also randomly sample one flow from each bin and compose additional Flow Sequence samples for training.

In the implementation of DoLLM, we set the number of bins to 64 and configure the training set to generate a sufficient total of 15,000 Flow Sequences. Incidentally, the described method of generating Flow Sequences not only allows LLMs to better uncover the correlations between flows but also acts as a stratified sampling of the network globally. Therefore, we conjecture that this approach also enables LLMs to better capture the overall state of the network.

\subsection{Flow Tokenizer}
\label{design_Flow_Tokenizer}

\begin{figure*}[htb]
\vspace{-2mm}
    \centering
    \begin{minipage}{0.33\textwidth}
        \includegraphics[width=0.9\linewidth]{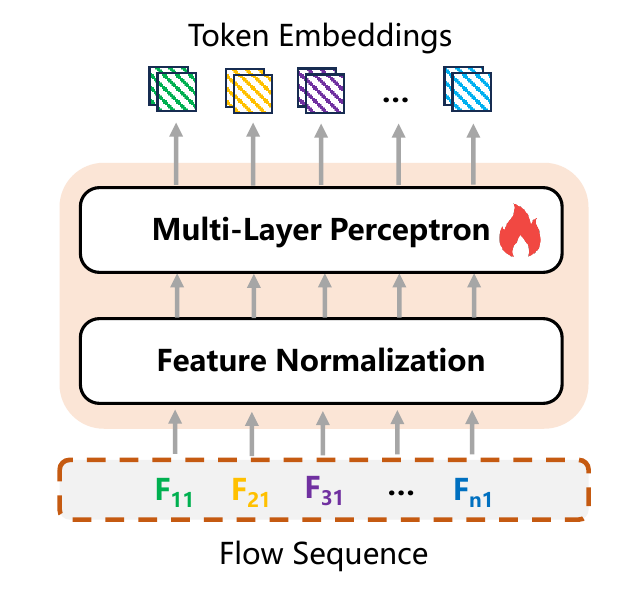} 
        \caption{Illustration of the workflow for Flow Tokenizer.}
        \label{Flow_Tokenizer}
    \end{minipage}
    \hfill
    \begin{minipage}{0.32\textwidth}
        \includegraphics[width=\linewidth]{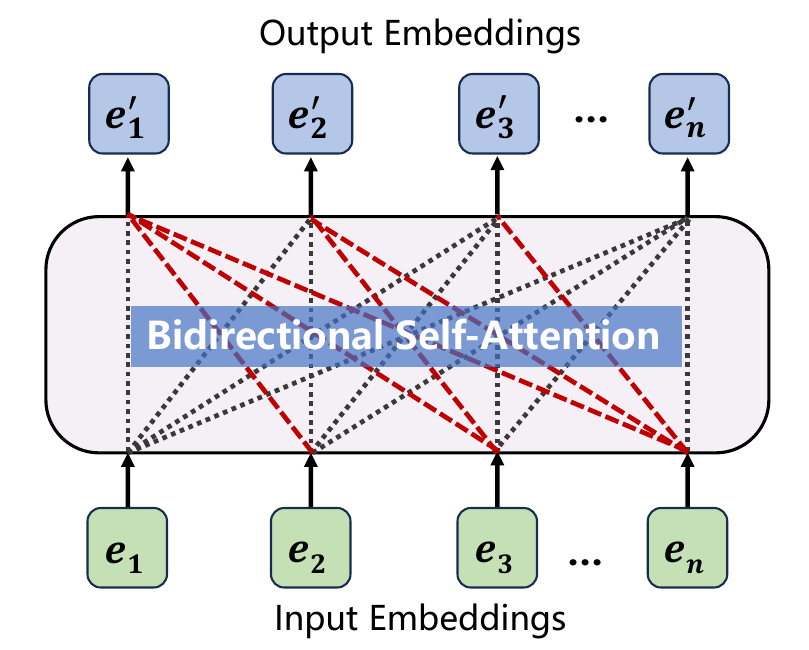} 
        \vspace{0mm}
        \caption{The self-attention mechanism in LLMs after removing the causal mask. }
        \label{self attention}
        
    \end{minipage}
    \hfill
    \begin{minipage}{0.32\textwidth}
        \includegraphics[width=0.95\linewidth]{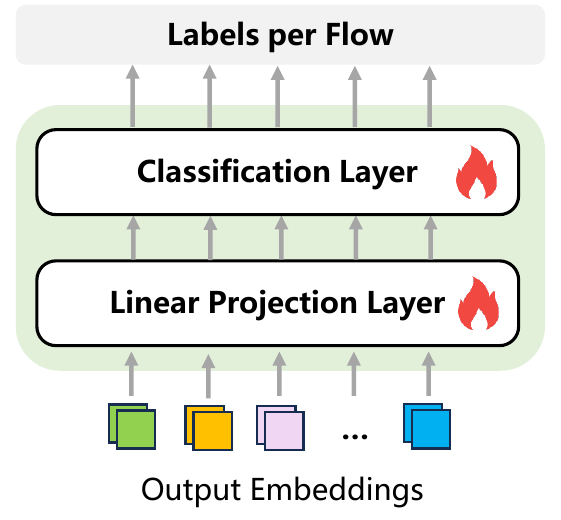} 
        \caption{Illustration of the workflow for Classification Projection.}
        \label{class_projection}
    \end{minipage}
\end{figure*}

Flow is a modality different from text, which LLMs cannot directly process. Therefore, \textbf{we designed the Flow Tokenizer to encode flow information into token embeddings within the LLMs' semantic space, allowing the LLMs to process them.} The architecture of Flow Tokenizer is shown in Figure ~\ref{Flow_Tokenizer}, which consists of two parts: Feature Normalization and a Multi-Layer Perceptron (MLP). In this section, we first describe how to select features to describe a flow, and then we introduce each of the two modules in the Flow Tokenizer.

The features used to describe flows can be divided into two types: {\it categorical features} and {\it numerical features}. The former represents categories without any numerical significance, while the latter has explicit numerical meanings. For categorical features, we select \textit{src\_port}, \textit{dst\_port}, and \textit{proto}. These attributes effectively reflect the flow's protocol and upper-layer applications, which are closely linked to attack vectors. It should be noted that we do not use source and destination IP addresses as features. This is because Carpet Bombing attacks do not have a fixed target, resulting in the randomness of destination IP addresses, which makes them difficult to use as a basis for detection. Also because of IP Spoofing, the source IP becomes unreliable. For numerical features, we select \textit{total\_bytes}, \textit{total\_pkts}, \textit{mean\_pkt\_len}, \textit{max\_pkt\_len}, \textit{min\_pkt\_len}, and \textit{std\_pkt\_len}. These six features can reflect the flow rate and the variability in packet size, effectively characterizing the behavior of the flow.

The role of Feature Normalization is to normalize the aforementioned features, generating a feature vector for each flow. For numerical features, we perform Min-max normalization, while for categorical features, we have to make them quantifiable, so we use frequency encoding, where the frequency of occurrence of these categories replaces the original feature values. This method is a common data preprocessing technique in machine learning and aligns well with our DDoS detection scenario, as the frequent occurrence of certain protocols or ports often signifies abnormal behavior. The above normalization process is shown in Equation ~\ref{norm}. Here, \(count(x)\) represents the number of occurrences of a specific feature value within its category, and \(F\) represents the total number of flows. After normalization, the nine features will compose the feature vector of the flow.

\vspace{-2mm}
\begin{equation}
\label{norm}
    x\_norm = 
\begin{cases} 
\frac{x - \min(x)}{\max(x) - \min(x)}, & \text{if } x \text{ is numeric feature} \\
\\
\frac{count(x)}{F}, & \text{if } x \text{ is categorical feature} \\
\end{cases}
\end{equation}

The function of MLP is to map the feature vectors into the semantic space of LLMs, achieving modal alignment between flows and text. In the architecture of LLMs, each token needs to be converted into a high-dimensional token embedding (e.g. 4096D in the Llama2) for processing. To this end, we designed a two-layer MLP to transform the feature vectors of each flow in the Flow Sequence into high-dimensional token embedding, enabling LLMs to handle them.

\subsection{Bidirectional Self-Attention LLMs}
\label{design_llm}
We use Llama2-7B~\cite{touvron2023llama} as the backbone, which is currently one of the most popular open-source LLMs. All Flows in the Flow Sequence, encoded by the Flow Tokenizer, are processed through the backbone for inference. We extract embeddings from the last hidden state output of backbone as the representation for each flow.

However, directly using LLMs for representation learning of flows still presents limitations, which arise from the masked self-attention~\cite{transformer} mechanism in LLMs. Most existing LLMs are based on a Transformer's decoder-only architecture, designed for text generation. Since only previously generated tokens can be accessed during the generation of new tokens, a Causal Mask is introduced that limits the self-attention mechanism. This Causal Mask makes the self-attention mechanism single-directional, allowing each token to only establish connections with previously existing tokens. This single-direction self-attention is disadvantageous for tasks like Token classification, which require representation learning of tokens because it prevents establishing connections between any two tokens, thereby losing a lot of information~\cite{tokenclass_li2023label}. Therefore, 
we have removed the causal mask in Llama2.

After removing the causal mask, the self-attention mechanism of LLMs is depicted in Figure ~\ref{self attention}. The black lines represent the original single-directional self-attention in LLMs, while the red lines indicate the attention pairs added after the removal of the causal mask.  \textbf{LLMs transition to Bidirectional self-attention, enabling the establishment of connections between each pair of flows in the Flow Sequence and uncovering the correlations between flows.} This improves the capability to extract representations of flows. This not only helps in identifying clusters of attack flows with similar characteristics but also effectively captures the differences between attack flows and normal flows, exposing the disguises of attack flows.

\subsection{Classification Projection}
\label{design_class}
\textbf{The Classification Projection evaluates the embedding of each flow from the Flow Sequence outputted by the LLMs to determine its label.} It performs binary classification on the embedding of each flow, sequentially outputting the probabilities that each flow is either an attack or a benign flow, and makes a determination based on the relative size of these probabilities. 

The classification Projection consists of two layers: a Linear Projection Layer and a Classification Layer. The Linear Projection Layer is a simple linear layer, whose primary function is to project the high-dimensional embeddings outputted by the LLMs into a lower-dimensional vector space (set to 256D). Given the high dimensionality of the LLM's output, which can include noise, reducing the dimensions through this linear layer allows for more effective extraction of useful information, thereby enhancing performance in classification tasks. The Classification Layer is another linear layer that performs binary classification based on the reduced-dimension embeddings from the Linear Projection Layer, outputting the probability that each flow is an attack flow.

\subsection{Put Them Together}
In this subsection, we will put the module toghther and summarize the inference and training pipeline of DoLLM.

During the inference process, at regular intervals, the Flow Collection Device will pass a batch of flows to DoLLM for detection. These flows first go through the Flow Sequentializer to generate  Flow Sequences. The Flow Sequences are then converted into Token Embeddings by the Flow Tokenizer and representations are subsequently extracted using the LLMs backbone. Finally, they pass through the Classification Projection to generate the probability of each Flow in the Flow Sequence being an attack flows. 

During the training process, DoLLM generates a sufficient amount of Flow Sequence training samples according to the procedure outlined in \S~\ref{design_Flow_Sequence_Generator}. These samples are then passed forward through the described inference process to obtain the predicted probability for each Flow in the Flow Sequence. Afterwards, the loss is computed using cross-entropy as the loss function, and back-propagation is performed. 

We use Equ.~\ref{cross-entropy} as the loss function.
Here, \( N \) represents the length of the Flow Sequence, which equal to the number of bins. \( y_i \) denotes the true label of the \( i \)-th flow, where 0 indicates a benign flow and 1 indicates an attack flow. \( \hat{y}_i \) represents the probability that the \( i \)-th flow is part of an attack. During this process, the network parameters of the LLM Backbone remain frozen and are not optimized, while the parameters of the Flow Tokenizer and Classification Projection are updated.

\begin{equation}
    L = -\frac{1}{N} \sum_{i=1}^{N} \left[ y_i \log(\hat{y}_i) + (1 - y_i) \log(1 - \hat{y}_i) \right]
    \label{cross-entropy}
\end{equation}
\section{Implementation}
\label{impl}

Because Carpet Bombing attacks target the access links of network infrastructures like IDCs, ISPs typically have management authority over these links. Therefore, when detection devices suspect a Carpet Bombing attack, they usually divert traffic on the access link to scrubbing devices for packet-level filtering. However, since current detection mechanisms are generally rule-based and cannot precisely identify specific Carpet Bombing attack flows, they resort to coarse-grained, large-scale diversion, 
even resulting in diversion across entire network segments. 
Among this traffic, there may be a significant amount of benign flow. This can cause considerable additional latency and impose a substantial overhead on the cleaning devices. 

\begin{figure}
\vspace{-2mm}
    \centering
    \includegraphics[width=0.85\linewidth]{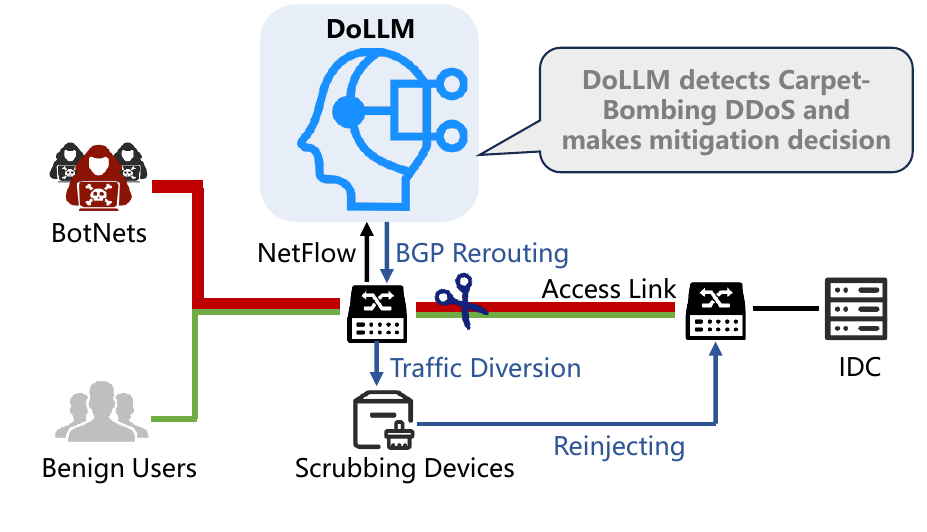}
    \vspace{-2mm}
    \caption{DoLLM system deployment in a Real ISP.}
    \label{Deployment}
    \vspace{-2mm}
\end{figure}

To address this, \textbf{we designed a practical deployment plan for DoLLM targeting a top-3 countrywide ISP.} The system primarily consists of a decision-making system based on DoLLM and scrubbing devices. The system architecture is illustrated in Figure ~\ref{Deployment}. Considering the immense volume of internet traffic in practice, we conduct flow-level detection based on NetFlow~\cite{netflow_claise2004cisco} whose sampling ratio can be set to 4000:1, and the Reports are sent to the DoLLM every minute.

DoLLM performs flow-level detection based on the NetFlow reports, identifying attack flows. Subsequently, DoLLM summarizes key information such as the five-tuple of attack flows, the traffic volume of attack flows, and Top N attack source IPs, among other alert data. Based on these alerts, diversion strategies are formulated and BGP Rerouting Announcement are made. The suspicious traffic is then diverted to scrubbing devices for packet-level filtering, and the cleaned traffic is reinjected into the link. It's important to note that the process of deciding on diversion strategies and the rules for scrubbing devices are not the focus of this paper. We believe that DoLLM's capability for detailed per-flow detection can provide more precise alert information, enabling finer-grained traffic diversion and packet-level scrubbing.

In practical deployment, there are still many challenges, such as how to train and the complexity of the model. For the first, DoLLM can be trained offline. Under lawful conditions, a large amount of traffic within the ISP's network can be collected as background benign traffic, and combined with attack flow information generated in the laboratory to create a dataset for training. This approach not only avoids harming actual network but also allows for the design of a variety of attack scenarios to enhance the model's generalization. For the latter, a method based on traffic thresholds can be designed to trigger DoLLM, ensuring that DoLLM is activated only in flooding situations. Additionally, deploying multiple DoLLM units in a distributed manner can reduce the load on each device, enhancing overall system efficiency and scalability.

\section{Evaluation}
\label{eva}
In this section, we will evaluate the performance of DoLLM in Carpet Bombing detection through experimental evaluation, using both the CIC-DDoS2019~\cite{cic_sharafaldin2019developing} dataset and real-world NetFlow reports collected from a top-3 countrywide ISP during Carpet Bombing attack events.

\subsection{Experiments Settings}
\label{setting}
In this subsection, we will describe the configurations of the DoLLM and the dataset.

\textbf{Configurations of DoLLM.} We implement DoLLM using Pytorch and download the open-source \textit{Llama2-7b-chat} model parameters from \textit{Huggingface}\footnote{The Hugging Face website offers a hub for exploring and implementing cutting-edge AI models and tools, \textit{https://huggingface.co}.}. We trained and tested DoLLM on a Linux server equipped with an AMD Ryzen 9 7950X 16-Core CPU and an NVIDIA GeForce RTX 4090 GPU. To enable the \textit{Llama2-7b-chat} model to run on the GPU, we set the floating-point precision of Llama2-7b to \textit{bfloat16}.

\textbf{CIC-DDoS2019 based Evaluation.} CIC-DDoS2019~\cite{cic_sharafaldin2019developing} dataset is a benchmark in the field of DDoS detection, containing 11 types of DDoS attacks, such as NTP flood and DNS flood, and provides flow-level statistical information. However, there are some issues with this dataset. Firstly, it only covers traditional DDoS attacks targeted at individual victim IPs. Secondly, the amount of normal traffic data is insufficient, making it difficult to effectively evaluate model performance. To address these issues, we use backbone network traffic collected from the MAWI~\cite{mawi_kenjiro2000traffic} dataset (January 1, 2024) as benign background traffic and mix it with the attack flows from CIC-DDoS2019 to create a Carpet Bombing dataset. We start by using the five-tuples to calculate statistics for each flow in the MAWI dataset. Then, we randomly select some flows from both datasets to mix, assigning random timestamps to each flow to simulate the characteristic of random arrival. By adjusting the mix ratio and the composition of attack types, we simulate various scenarios. Additionally, to mimic the multiple victim IPs and IP spoofing seen in Carpet Bombing, we modify the destination IPs to IPs within the range of 10 Class C network segments (totaling 2560 IPs) randomly, and change the source IPs to random numbers. 


\textbf{Real ISP Trace based Evaluation.} We also tested the performance of DoLLM using traces from another real Carpet Bombing incident that occurred within a top-3 countrywide ISP. This Carpet Bombing event happened in November 2023, targeting a public Cloud. The attack triggered network security device alerts, enabling us to capture the attack flow information in the cleaning devices. The attack flow data was formed using a NetFlow 4000:1 sampling ratio, with the attack consisting of multiple vectors using a mix of TCP, UDP, and ICMP protocols. Due to constraints, we only extracted NetFlow information for 1222 attack flows. We also extracted NetFlow information for benign background traffic from the same ISP, gathered during the time period adjacent to the attack using the same sampling ratio. We mixed 12,220 benign flows with the attack flows and assigned random timestamps to create a dataset.  During this process, sensitive information such as IP addresses was encrypted, and since NetFlow data does not contain any packet payload information, no personal privacy was compromised.

\textbf{Performance Comparison.} We have selected Extreme Gradient Boosting (XGBoost)~\cite{chen2016xgboost}, Support Vector Machine (SVM), and Multi-Layer Perceptron (MLP) as comparison methods for DoLLM. XGBoost and SVM are regarded as representatives of machine learning models based on explicit features, while MLP is a representative of deep learning. All comparison methods use the same flow features and normalization methods as DoLLM for classification. Each dataset is divided into training, validation, and testing sets in a 6:2:2 ratio, organized by timestamps. All methods will use the validation set to select the model that achieves the best F1 Score for testing on the testing set. For DoLLM and MLP, we conduct training for 20 rounds and choose the model that achieves the highest F1 score on the validation set. For XGBoost and SVM, we use grid search to find the parameters that yield the best F1 score on the validation set. For each method, we calculate its \textit{accuracy}, \textit{precision}, \textit{recall}, and \textit{F1 score} as performance metrics. Accuracy represents the proportion of correct predictions and is commonly used in balanced datasets. The F1 score measures the balance between precision and comprehensiveness in the detection of attack flows, making it suitable for use in imbalanced datasets.

\subsection{CIC-DDoS2019 Based Evaluation}
\label{cic-test}
In this subsection, we created three different Carpet Bombing datasets based on the CIC-DDoS2019 dataset, categorized as \textit{Overall Performance}, \textit{imbalanced classification}, and \textit{Zero-Shot} and tested the performance of DoLLM in these scenarios.

\textbf{Overall performance in a balanced dataset}, which is the most standard and straightforward setting in machine learning. It includes 20,000 attack flows and 20,000 benign flows. The test results are presented in Table ~\ref{table_overall}.

\begin{table}[t]
\centering
\caption{Overall performance in balanced classification}
\label{table_overall}
\begin{tabular}{@{}ccccc@{}} 
\toprule
 Methods & Accuracy & Precision & Recall & F1 Score \\
\midrule
DoLLM   & \textbf{0.982} & 0.983 & 0.982 & \textbf{0.983} \\
XGBoost & 0.968 & \textbf{0.985} & 0.951 & 0.967 \\
MLP     & 0.979 & 0.966 & \textbf{0.994} & 0.980 \\
SVM     & 0.935 & 0.891 & 0.993 & 0.939 \\
\bottomrule
\end{tabular}
\end{table}

As can be seen in the balanced dataset, DoLLM achieves the highest accuracy and F1 score, indicating that it has the best detection performance, capable of correctly classifying flows while accurately and comprehensively identifying attack flows. It should be noted that the balanced dataset represents a relatively simple scenario, where the other three methods also perform well. This is because, in a balanced data environment, machine learning models based on hand-crafted feature engineering can utilize sufficient positive and negative samples to delineate the differences between them, although there is still a gap compared to DoLLM.

\begin{figure}[t]
    \vspace{0mm}
    \centering
    \begin{subfigure}[b]{0.48\linewidth} 
        \centering
        \includegraphics[width=\linewidth]{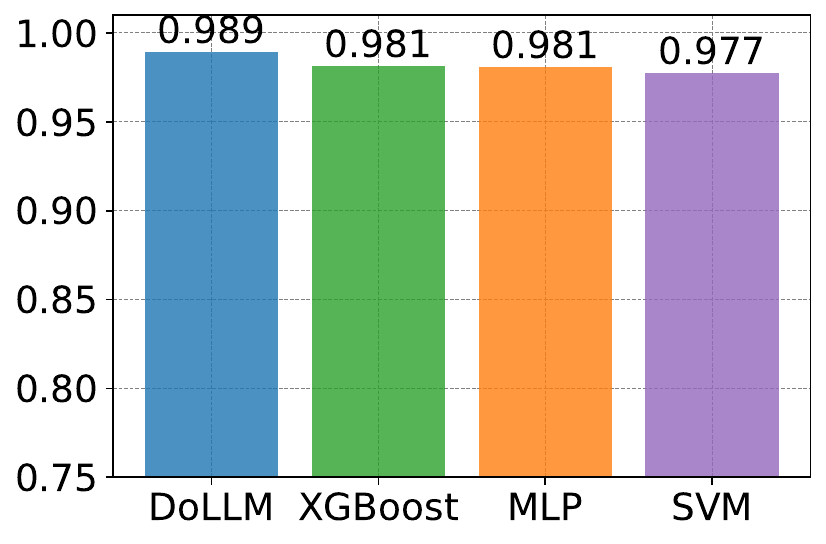} 
        \caption{Accuracy}
    \end{subfigure}
    \hfill
    \begin{subfigure}[b]{0.48\linewidth}
        \centering
        \includegraphics[width=\linewidth]{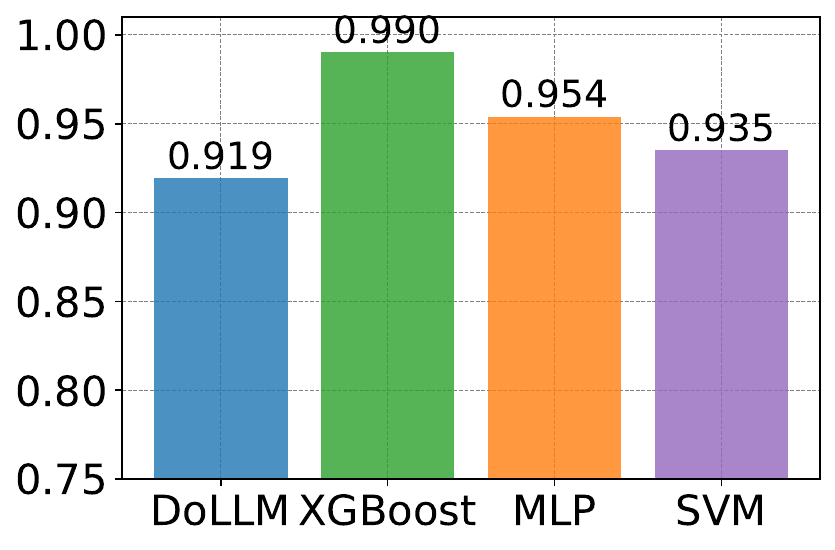}
        \caption{Precision}
    \end{subfigure}

    \vspace{1ex} 

    \begin{subfigure}[b]{0.48\linewidth}
        \centering
        \includegraphics[width=\linewidth]{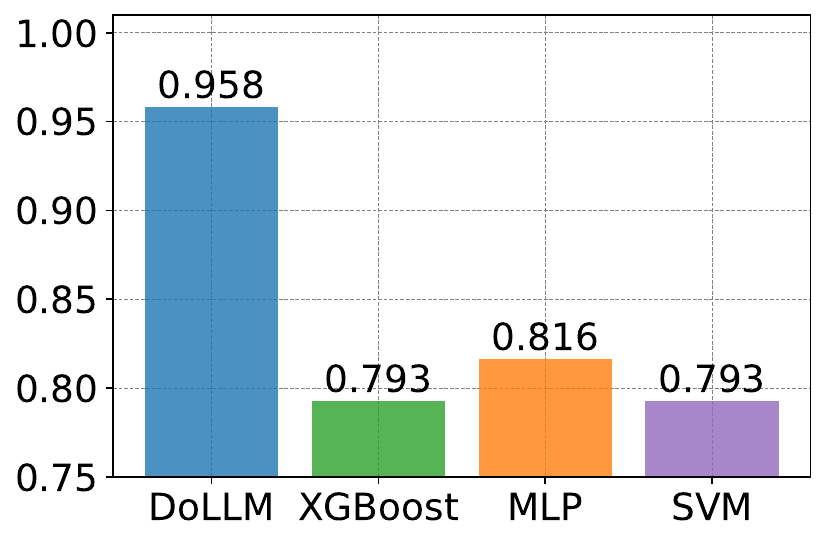}
        \caption{Recall}
    \end{subfigure}
    \hfill
    \begin{subfigure}[b]{0.48\linewidth}
        \centering
        \includegraphics[width=\linewidth]{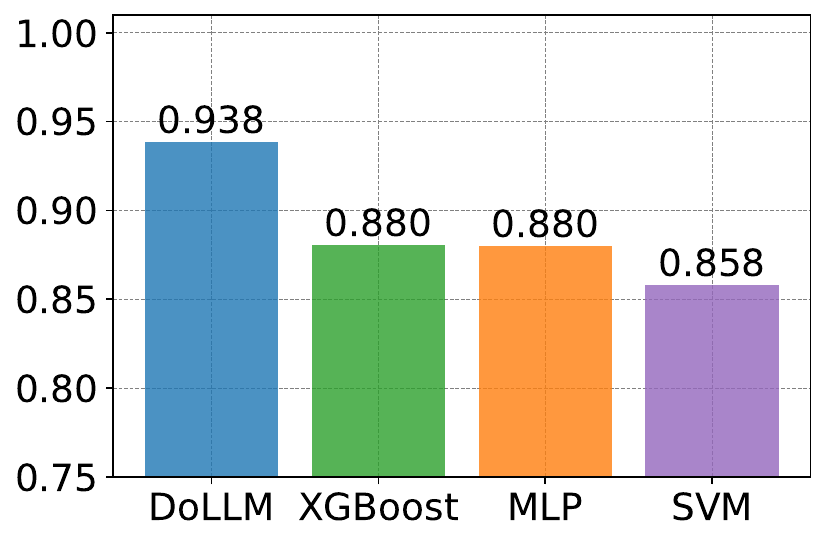}
        \caption{F1 Score}
    \end{subfigure}

    \caption{Performance metrics for imbalanced classification.}
    \label{fig:imbalance}
    \vspace{-4mm}
\end{figure}

\textbf{Performance in imbalance classification.} In this scenario, the number of benign flows significantly exceeds the number of attack flows. We believe this better reflects the actual network environment since Carpet Bombing is a rare event, meaning the number of attack flows we can collect is much smaller than the number of benign flows available. We created an imbalanced dataset for testing by mixing 2,000 attack flows with 20,000 benign flows. The test results are displayed in Figure ~\ref{fig:imbalance}.

In the context of imbalanced classification, DoLLM outshines other methods by securing the highest marks in accuracy, recall, and F1 score, underscoring its superior capability to detect attacks. Notably, the F1 score, a crucial metric for imbalanced datasets, is 6.6\% higher for DoLLM compared to other methods, providing a more balanced assessment of its classification prowess. This substantial improvement highlights DoLLM's precision and thoroughness in identifying attack flows, which is especially effective against complex threats like Carpet Bombing. Additionally, DoLLM's recall exceeds that of other algorithms by at least 17.4\%, affirming its ability to capture most attack flows and enhance the detection of intricate attack strategies.

The observed differences primarily stem from the fact that comparative methods only consider features of individual flows. In imbalanced datasets, where attack flows are scarce, these methods fail to learn the essential characteristics of attack flows, thus tending to classify them as benign flows. Consequently, these methods cannot effectively detect attack flows. In contrast, DoLLM accounts for the correlations between flows. It first learns the inherent characteristics of flows at various positions within the Flow Sequence, and secondly, through self-attention, it effectively captures differences between flows, serving as an additional supervisory signal. Therefore, even in the presence of sample imbalance, DoLLM can achieve good results. This also demonstrates that there is no need to deliberately create balanced datasets through preprocessing when training DoLLM.

\begin{figure}[t]
    \vspace{0mm}
    \centering
    \begin{subfigure}[b]{0.48\linewidth} 
        \centering
        \includegraphics[width=\linewidth]{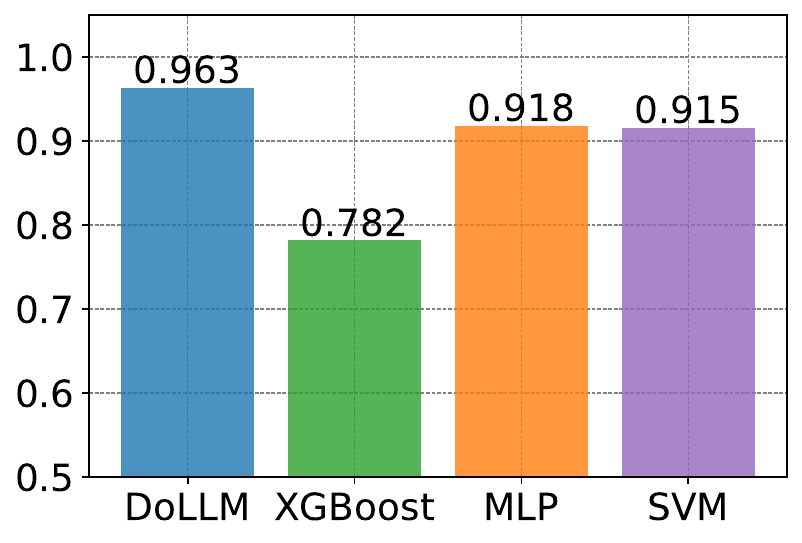} 
        \caption{Accuracy}
    \end{subfigure}
    \hfill
    \begin{subfigure}[b]{0.48\linewidth}
        \centering
        \includegraphics[width=\linewidth]{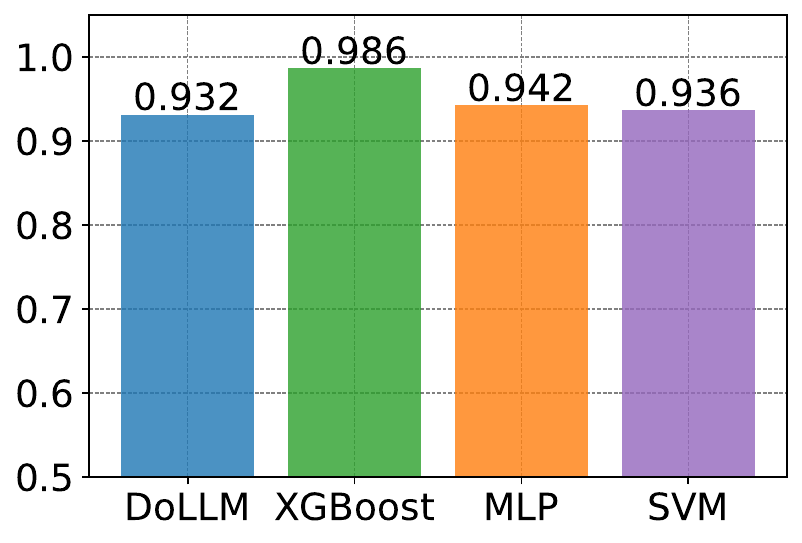}
        \caption{Precision}
    \end{subfigure}

    \vspace{1ex} 

    \begin{subfigure}[b]{0.48\linewidth}
        \centering
        \includegraphics[width=\linewidth]{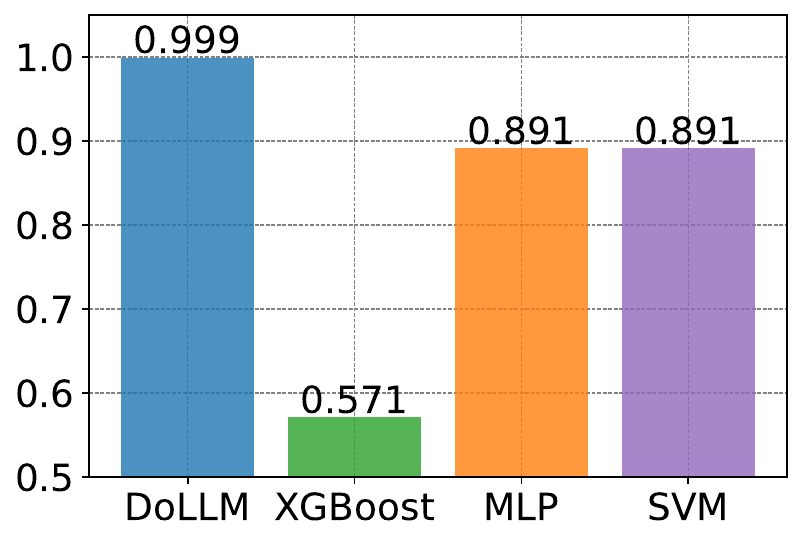}
        \caption{Recall}
    \end{subfigure}
    \hfill
    \begin{subfigure}[b]{0.48\linewidth}
        \centering
        \includegraphics[width=\linewidth]{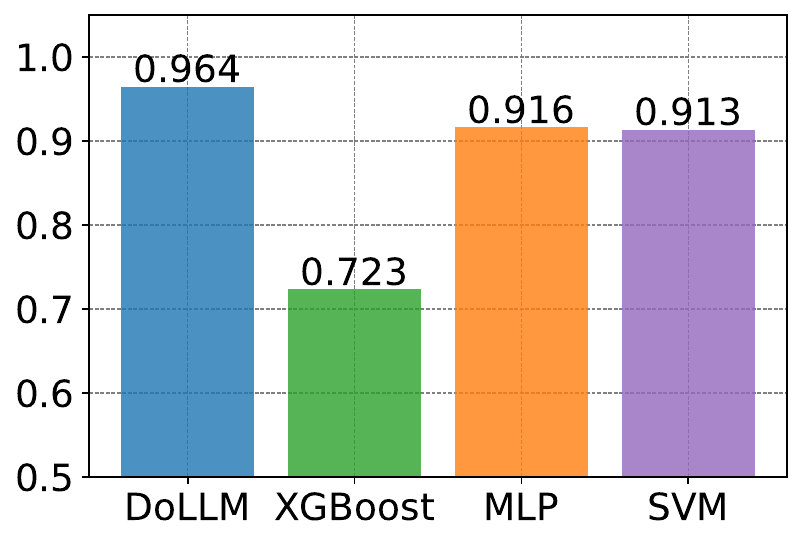}
        \caption{F1 Score}
    \end{subfigure}

    \caption{Performance metrics for zero-shot.}
    \label{fig:zeroshot}
    \vspace{-4mm}
\end{figure}

\textbf{Performance in zero-shot.} One of our expectations for a DDoS detection model is its ability to discover new types of attacks, meaning it should have zero-shot capability. To facilitate this, we created two Carpet Bombing datasets with  different attack vectors, Dataset A and Dataset B. In Dataset A, we mixed three common attack vectors, DNS, NTP, and SYN flood, with benign flows from the MAWI dataset in a 20,000 : 20,000 ratio. In Dataset B, we mixed the remaining eight attack vectors from CIC-DDoS2019 with different benign flows at the same ratio. We trained the model on Dataset A and tested it on Dataset B. The test results are shown in Figure ~\ref{fig:zeroshot}. Additionally, we compared the differences between zero-shot and non-zero-shot scenarios. In the non-zero-shot scenario, we train on the training set of Dataset A and test on the testing set of Dataset A. In the zero-shot scenario, we train on the training set of Dataset A but test on Dataset B. The results are displayed in Table ~\ref{non-zero-shot and zero-shot}.

In the zero-shot scenario, DoLLM exhibits the best accuracy, recall, and F1 score, with accuracy and F1 score at least 4.9\% and 5.2\% higher than other algorithms, respectively. This indicates that DoLLM has a strong capability to detect new types of attacks. Furthermore, by comparing the non-zero-shot and zero-shot scenarios, it is evident that DoLLM not only achieves the best results in both scenarios but also shows the smallest performance decline. In zero-shot, DoLLM's accuracy only decreases by 1.6\% compared to non-zero-shot, while the best-performing other method, MLP, drops by 5.2\%, and XGBoost drops by as much as 19.1\%. This demonstrates that DoLLM has the best generalization ability, able to more effectively learn useful information from existing training data and apply it to new scenarios.

The primary reason for the performance decline of comparative methods in zero-shot scenarios is their excessive reliance on the specific features of each flow. These methods may struggle to adapt when features of new types of attacks change. However, new attacks may not be entirely different from existing ones. For instance, they might exhibit similar traffic characteristics but differ in the ports and protocols used. For such new types of attacks, DoLLM might utilize binning to position them in the same location within the Flow Sequence, thereby better leveraging previously learned experiences. Furthermore, DoLLM employs self-attention to more effectively focus on the differences between attack flows and other flows, enabling it to make more accurate judgments.

\begin{table}[t]
\vspace{-2mm}
\centering
\begin{tabular}{c|cc|cc}
\toprule
Methods & \multicolumn{2}{c|}{\parbox{2cm}{\centering non-zero-shot \\ ($A \to A$)}} & \multicolumn{2}{c}{\parbox{2cm}{\centering zero-shot \\ ($A \to B$)}} \\
\midrule
Indicator        & Accuracy & F1 score & Accuracy & F1 score \\
\midrule
DoLLM            & \textbf{0.979}     & \textbf{0.979} & \textbf{0.963}     & \textbf{0.964} \\
XGBoost          & 0.967     & 0.967 & 0.782     & 0.723 \\
MLP              & 0.969     & 0.970 & 0.918     & 0.916 \\
SVM              & 0.906     & 0.914 & 0.915     & 0.913 \\
\bottomrule
\end{tabular}
\caption{Comparison of performance across methods in non-zero-shot and zero-shot scenarios}
\label{non-zero-shot and zero-shot}
\end{table}

\subsection{Ablation and Parameter Experiments}
\begin{figure}[t]
\vspace{-4mm}
    \centering
    \includegraphics[width=0.8\linewidth]{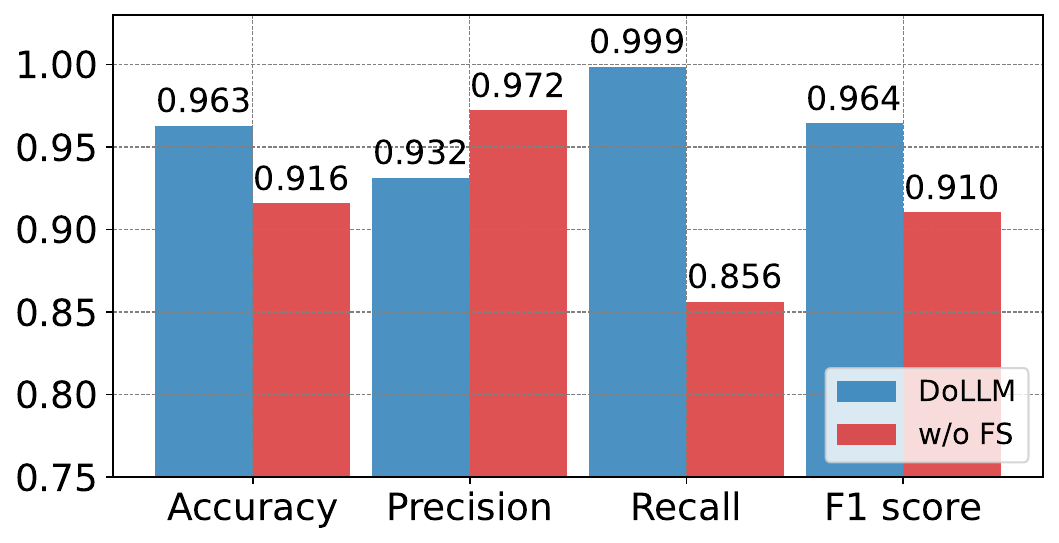}
    \caption{Performance Comparison between DoLLM and DoLLM without Flow Sequentializer (w/o FS).}
    \label{FSG}
    \vspace{-2mm}
\end{figure}

\textbf{Ablation Experiment on the Flow Sequentializer.} We conducted an experiment to assess the impact of the Flow Sequentializer in DoLLM. We removed the Flow Sequentializer module from DoLLM and instead generated Flow Sequence samples using a sliding window approach with a length of 64 and a stride of 1. In this setup, the flows in the Flow Sequence are arranged according to their temporal order of arrival. We tested the performance of DoLLM with and without the Flow Sequentializer module in the zero-shot scenario described in \S~\ref{cic-test}. The experimental results are shown in Figure ~\ref{FSG}.

It is evident that removing the Flow Sequentializer results in a significant decrease in the accuracy, recall, and F1 score of DoLLM. Specifically, accuracy and F1 decreased by 4.9\% and 5.6\% respectively, while recall saw a steep decline of 14.3\%. This indicates a substantial drop in DoLLM's detection capabilities after removing the Flow Sequentializer.

This decrease in performance is due to the lack of inherent temporal correlations between flows, meaning that Flow Sequences created in this manner lack consistent patterns. As a result, LLMs struggle to accurately learn representations for each flow. On the other hand, the Flow Sequentializer creates ordered Flow Sequences where each position’s flow has a similar meaning, effectively setting a 'grammar' for the Flow Sequence. This allows LLMs to better extract inter-flow correlations, thereby enhancing detection performance.

\begin{figure}[t]
\vspace{-4mm}
    \centering
    \includegraphics[width=0.9\linewidth]{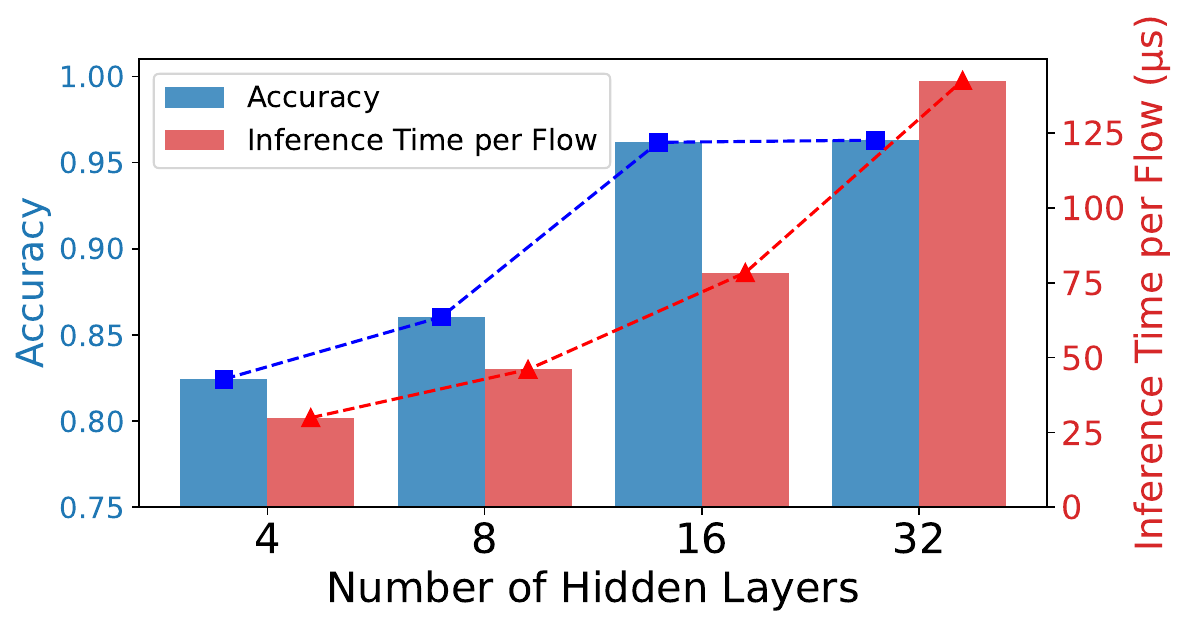}
    \caption{Accuracy and inference time per Flow of DoLLM with the number of Hidden Layers.}
    \label{hidden_layers}
    \vspace{-3mm}
\end{figure}

\textbf{Performance varies with the number of hidden layers.} In LLMs, the term "hidden layer" refers to the number of Transformer blocks within the model. In Llama2, the default number is 32. We varied the number of hidden layers and tested the performance of DoLLM in the zero-shot scenario described in \S~\ref{cic-test}. The test results are shown in Figure ~\ref{hidden_layers}.

Doubling the number of hidden layers in LLMs significantly increases computational demands and inference time, but does not necessarily enhance accuracy beyond a certain point. For example, 32 hidden layers offer nearly the same accuracy as 16. Since only a few layers may be needed to extract basic semantics, reducing the number of layers could balance detection performance and speed in practical applications.

\subsection{Real ISP Trace based Evaluation} 

\begin{figure}[t]
    \centering
    \begin{subfigure}[b]{0.48\linewidth} 
        \centering
        \includegraphics[width=\linewidth]{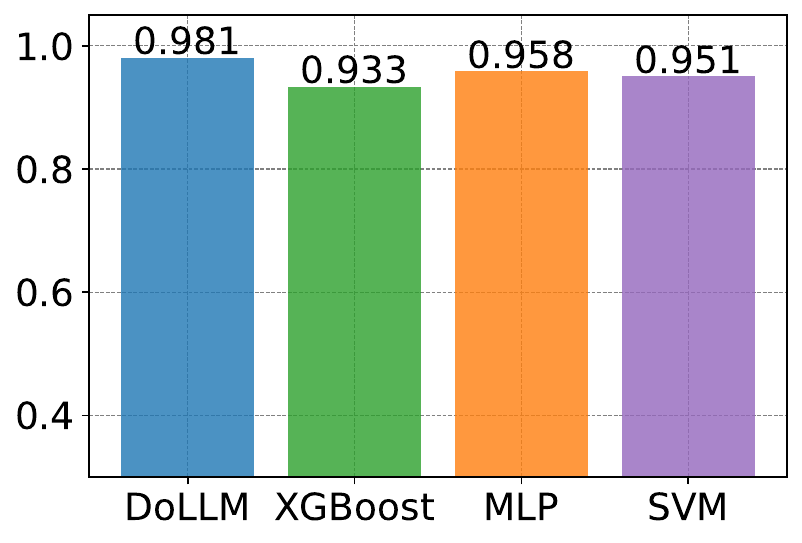} 
        \caption{Accuracy}
    \end{subfigure}
    \hfill
    \begin{subfigure}[b]{0.48\linewidth}
        \centering
        \includegraphics[width=\linewidth]{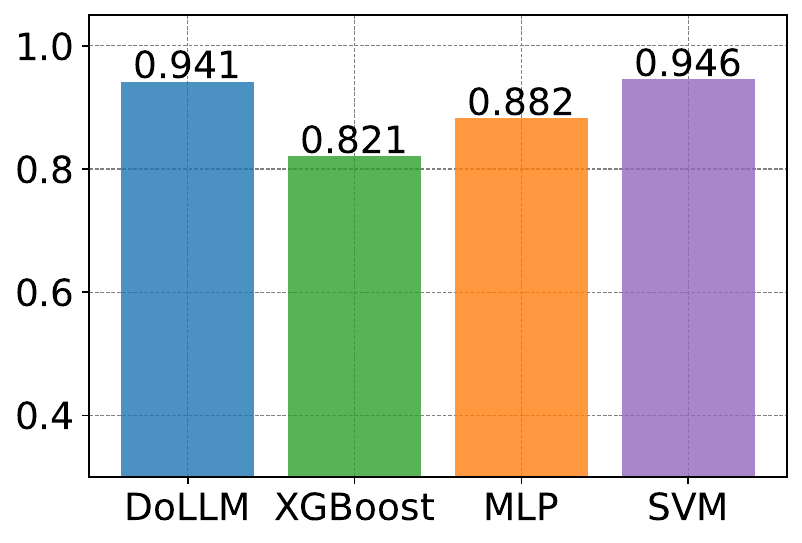}
        \caption{Precision}
    \end{subfigure}

    \vspace{1ex} 

    \begin{subfigure}[b]{0.48\linewidth}
        \centering
        \includegraphics[width=\linewidth]{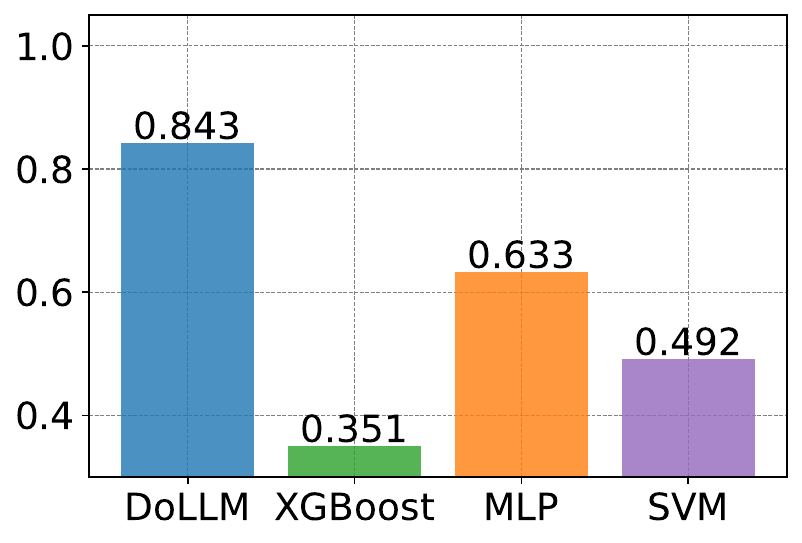}
        \caption{Recall}
    \end{subfigure}
    \hfill
    \begin{subfigure}[b]{0.48\linewidth}
        \centering
        \includegraphics[width=\linewidth]{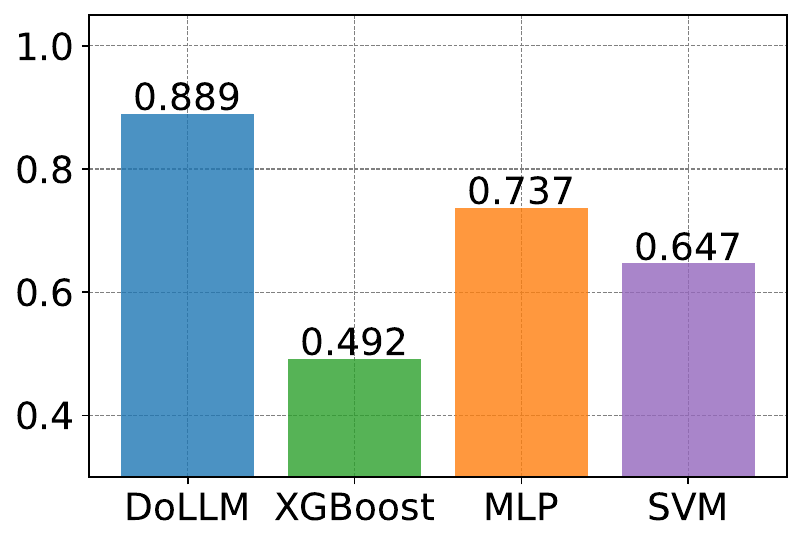}
        \caption{F1 Score}
    \end{subfigure}

    \caption{Performance metrics for Real ISP Trace.}
    \label{fig:netflow}
\end{figure}

We conducted tests using the Carpet Bombing dataset created from real ISP trace as described in \S~\ref{setting}. Since NetFlow doesn't record \textit{max\_pkt\_len}, \textit{min\_pkt\_len}, or \textit{std\_pkt\_len}, we used only the remaining six features mentioned in \S~\ref{design_Flow_Tokenizer} for classification. The results are shown in Figure ~\ref{fig:netflow}.

DoLLM has the highest accuracy, recall, and F1 in real ISP Trace, and its Recall and F1 values far exceed those of the remaining methods. The recall was at least 33.2\% higher than other methods, and the F1 score improved by at least 20.6\%. This is because, under high sampling ratios, the granularity of flow information collected by NetFlow is coarse, making the features of attack flows and benign flows overly similar. Machine learning methods based on hand-crafted feature engineering cannot learn the differences between them, thus failing to make correct classifications. However, DoLLM can capture the subtle differences between flows through its self-attention mechanism, improving classification performance.

In conclusion, even under conditions of coarse-grained flow information from high ratio NetFlow sampling, DoLLM can still detect Carpet Bombing both comprehensively and precisely. Therefore, DoLLM has the potential to be practically deployed in ISPs.

\section{Limitations}
\label{limi}
In this section, we will discuss some potential limitations of DoLLM, provide analysis and suggest possible solutions.

\textbf{Higher Computational Overhead.} Because of the huge network parameters of LLMs, DoLLM experiences significant computational overhead and inference latency, which limits its effectiveness in real-time detection. However, we believe that this is a common challenge when exploring the use of LLMs in latency-sensitive applications, such as autonomous driving~\cite{autodriving_cui2024survey} or robotics~\cite{robotic_zeng2023large}. Addressing this issue relies heavily on advancements in model compression for LLMs~\cite{compression_zhu2023survey}. Nevertheless, we have made efforts to improve inference efficiency. Firstly, by modeling DDoS detection as a Token classification problem, LLMs can output representations for multiple flows with each inference. This approach, compared to using complex textual descriptions of the network environment to directly output detection results from LLMs~\cite{llmddos_guastalla2023application}, significantly improves detection efficiency. Secondly, our experiments have shown that reducing the number of hidden layers in LLMs can effectively lower inference latency while maintaining acceptable detection performance.

\textbf{No Prompt Engineering.} In DoLLM, we did not use textual prompts to input information into LLMs due to the modal gap between flow information and text. However, since LLMs have been pretrained with massive text corpora, appropriate prompts can activate LLMs' knowledge, thereby enhancing the quality of the output. This process is also known as Prompt Engineering~\cite{promptengineering_sahoo2024systematic}. Appropriate Prompt Engineering can also enhance the representation learning capabilities of open-source LLMs~\cite{time_zhou2024one}. In our future work, we plan to provide LLMs with additional network information through appropriate prompts, exploring potential improvements.

\section{Related Work}
\label{related}
With the advancement of the Transformer~\cite{transformer} architecture, pretrained models~\cite{pretrain_qiu2020pre} have become a crucial technology in natural language processing and computer vision, and they are also one of the core technologies for LLMs. Currently, there are efforts to employ pretrained models or LLMs to tackle various tasks within the field of network security.

\textbf{Pretrained model for network security.} YaTC~\cite{YaTC_zhao2023yet} and ET-BERT~\cite{ETBERT-lin2022bert}, based respectively on a masked autoencoder and BERT, pretrain models to extract representations from network traffic. These pretrained models are then fine-tuned for downstream traffic classification tasks using these representations. However, both models rely on packet payloads as inputs, which are challenging to handle in access links with massive data packets. Additionally, due to the limited scale of publicly available traffic datasets and the substantial computational resources required for pretraining, it is difficult for ordinary research teams to train high-quality models. This also drives us to apply the representational abilities of pretrained LLMs to DDoS detection.

\textbf{LLMs for network security.} Existing work largely employs the close-source GPT~\cite{achiam2023gpt} family of LLMs to accomplish certain tasks. Ziems et al.~\cite{llmsecurity_ziems2023explaining} used GPT-4 to explain the decision-making in network intrusion detection, enhancing the interpretability of detection mechanisms. Yan et al.~\cite{llmsecurity_yan2023prompt} used GPT-4 to generate descriptive texts, employed for representation learning of API sequences, facilitating dynamic analysis of malware. Guastalla et al.~\cite{llmddos_guastalla2023application} has employed GPT for few-shot and zero-shot detection of DDoS attacks. However, due to limitations of the training corpora, LLMs may not possess adequate knowledge of this tasks, potentially leading to poor problem-solving and the emergence of hallucinations. Therefore, we believe that using open-source LLMs for representation learning is a better approach than directly employing close-source LLMs for inference. \textbf{To our knowledge, we are the first to design a DDoS detection model using open-source LLMs as the backbone.}
\section{Conclusion}
\label{conclu}
In this paper, we first analyze and summarize the threats posed by Carpet Bombing, a new form of DDoS attack, as well as the challenges in defending against it. We then introduce DoLLM, the first DDoS detection model that utilizes open-source LLMs as backbone. By organizing temporally unordered flows into a structured sequence, DoLLM enables LLMs to explore the correlations between flows and effectively capture the representation of different flows, which is further used for Carpet Bombing detection. We evaluated the performance of DoLLM using public datasets and real ISP trace, demonstrating its effective detection capabilities against Carpet Bombing. Additionally, we have designed the deployment of DoLLM in actual ISPs.

\bibliographystyle{plain}
\bibliography{main}

\begin{thebibliography}{10}

\bibitem{achiam2023gpt}
Josh Achiam, Steven Adler, Sandhini Agarwal, Lama Ahmad, Ilge Akkaya, Florencia~Leoni Aleman, Diogo Almeida, Janko Altenschmidt, Sam Altman, Shyamal Anadkat, et~al.
\newblock Gpt-4 technical report.
\newblock {\em arXiv preprint arXiv:2303.08774}, 2023.

\bibitem{autoregressive_aiello2023jointly}
Emanuele Aiello, Lili Yu, Yixin Nie, Armen Aghajanyan, and Barlas Oguz.
\newblock Jointly training large autoregressive multimodal models.
\newblock {\em arXiv preprint arXiv:2309.15564}, 2023.

\bibitem{accturbo-alcoz2022aggregate}
Albert~Gran Alcoz, Martin Strohmeier, Vincent Lenders, and Laurent Vanbever.
\newblock Aggregate-based congestion control for pulse-wave ddos defense.
\newblock In {\em Proceedings of the ACM SIGCOMM 2022 Conference}, pages 693--706, 2022.

\bibitem{autoregressive_awadalla2023openflamingo}
Anas Awadalla, Irena Gao, Josh Gardner, Jack Hessel, Yusuf Hanafy, Wanrong Zhu, Kalyani Marathe, Yonatan Bitton, Samir Gadre, Shiori Sagawa, et~al.
\newblock Openflamingo: An open-source framework for training large autoregressive vision-language models.
\newblock {\em arXiv preprint arXiv:2308.01390}, 2023.

\bibitem{carpet_corero}
Emma Cadzow.
\newblock Carpet bomb ddos attacks: On the rise and evading detection.
\newblock \url{https://www.corero.com/threat-report-carpet-bomb-intro}.

\bibitem{chen2016xgboost}
Tianqi Chen and Carlos Guestrin.
\newblock Xgboost: A scalable tree boosting system.
\newblock In {\em Proceedings of the 22nd acm sigkdd international conference on knowledge discovery and data mining}, pages 785--794, 2016.

\bibitem{carpet_South_african}
Catalin Cimpanu.
\newblock 'carpet-bombing' ddos attack takes down south african isp for an entire day, 2019.
\newblock \url{https://www.zdnet.com/article/carpet-bombing-ddos-attack-takes-down-south-african-isp-for-an-entire-day/}.

\bibitem{netflow_claise2004cisco}
Benoit Claise.
\newblock Cisco systems netflow services export version 9.
\newblock Technical report, 2004.

\bibitem{autodriving_cui2024survey}
Can Cui, Yunsheng Ma, Xu~Cao, Wenqian Ye, Yang Zhou, Kaizhao Liang, Jintai Chen, Juanwu Lu, Zichong Yang, Kuei-Da Liao, et~al.
\newblock A survey on multimodal large language models for autonomous driving.
\newblock In {\em Proceedings of the IEEE/CVF Winter Conference on Applications of Computer Vision}, pages 958--979, 2024.

\bibitem{carpet_huawei}
China Telecom Cybersecurity Technology~Co. et~al.
\newblock Global ddos attack status and trend analysis in 2023, 2024.
\newblock \url{https://e.huawei.com/eu/material/networking/security/0c561b8fd2d342999cd402bcecf6d452}.

\bibitem{carpet_FastNetMon}
FastNetMon.
\newblock Rise of carpet bombing attacks, 2023.
\newblock \url{https://fastnetmon.com/2023/10/24/rise-of-carpet-bombing-ddos-attacks-and-ways-to-detect-and-defend-against-them-using-fastnetmon-advanced/}.

\bibitem{llmddos_guastalla2023application}
Michael Guastalla, Yiyi Li, Arvin Hekmati, and Bhaskar Krishnamachari.
\newblock Application of large language models to ddos attack detection.
\newblock In {\em International Conference on Security and Privacy in Cyber-Physical Systems and Smart Vehicles}, pages 83--99. Springer, 2023.

\bibitem{carpet_DRDoS}
Tiago Heinrich, Rafael~R Obelheiro, and Carlos~A Maziero.
\newblock New kids on the drdos block: Characterizing multiprotocol and carpet bombing attacks.
\newblock In {\em International Conference on Passive and Active Network Measurement}, pages 269--283. Springer, 2021.

\bibitem{time_jin2023time}
Ming Jin, Shiyu Wang, Lintao Ma, Zhixuan Chu, James~Y Zhang, Xiaoming Shi, Pin-Yu Chen, Yuxuan Liang, Yuan-Fang Li, Shirui Pan, et~al.
\newblock Time-llm: Time series forecasting by reprogramming large language models.
\newblock {\em arXiv preprint arXiv:2310.01728}, 2023.

\bibitem{link_kang2013crossfire}
Min~Suk Kang, Soo~Bum Lee, and Virgil~D Gligor.
\newblock The crossfire attack.
\newblock In {\em 2013 IEEE symposium on security and privacy}, pages 127--141. IEEE, 2013.

\bibitem{mawi_kenjiro2000traffic}
CHO Kenjiro.
\newblock Traffic data repository at the wide project.
\newblock In {\em Proc. USENIX 2000 Annual Technical Conference: FREENIX Track, San Diego, CA}, 2000.

\bibitem{carpet_netscout}
Richard~Hummel Kinjal~Patel.
\newblock Carpet-bombing, 2024.
\newblock \url{https://www.netscout.com/blog/asert/carpet-bombing}.

\bibitem{link_lee2013codef}
Soo~Bum Lee, Min~Suk Kang, and Virgil~D Gligor.
\newblock Codef: Collaborative defense against large-scale link-flooding attacks.
\newblock In {\em Proceedings of the ninth ACM conference on Emerging networking experiments and technologies}, pages 417--428, 2013.

\bibitem{Ner_li2020survey}
Jing Li, Aixin Sun, Jianglei Han, and Chenliang Li.
\newblock A survey on deep learning for named entity recognition.
\newblock {\em IEEE transactions on knowledge and data engineering}, 34(1):50--70, 2020.

\bibitem{image_li2023blip}
Junnan Li, Dongxu Li, Silvio Savarese, and Steven Hoi.
\newblock Blip-2: Bootstrapping language-image pre-training with frozen image encoders and large language models.
\newblock In {\em International conference on machine learning}, pages 19730--19742. PMLR, 2023.

\bibitem{video_li2023videochat}
KunChang Li, Yinan He, Yi~Wang, Yizhuo Li, Wenhai Wang, Ping Luo, Yali Wang, Limin Wang, and Yu~Qiao.
\newblock Videochat: Chat-centric video understanding.
\newblock {\em arXiv preprint arXiv:2305.06355}, 2023.

\bibitem{video_li2023llama}
Yanwei Li, Chengyao Wang, and Jiaya Jia.
\newblock Llama-vid: An image is worth 2 tokens in large language models.
\newblock {\em arXiv preprint arXiv:2311.17043}, 2023.

\bibitem{tokenclass_li2023label}
Zongxi Li, Xianming Li, Yuzhang Liu, Haoran Xie, Jing Li, Fu-lee Wang, Qing Li, and Xiaoqin Zhong.
\newblock Label supervised llama finetuning.
\newblock {\em arXiv preprint arXiv:2310.01208}, 2023.

\bibitem{link_liaskos2016novel}
Christos Liaskos, Vasileios Kotronis, and Xenofontas Dimitropoulos.
\newblock A novel framework for modeling and mitigating distributed link flooding attacks.
\newblock In {\em IEEE INFOCOM 2016-The 35th Annual IEEE International Conference on Computer Communications}, pages 1--9. IEEE, 2016.

\bibitem{ETBERT-lin2022bert}
Xinjie Lin, Gang Xiong, Gaopeng Gou, Zhen Li, Junzheng Shi, and Jing Yu.
\newblock Et-bert: A contextualized datagram representation with pre-training transformers for encrypted traffic classification.
\newblock In {\em Proceedings of the ACM Web Conference 2022}, pages 633--642, 2022.

\bibitem{image_liu2023llava}
Shilong Liu, Hao Cheng, Haotian Liu, Hao Zhang, Feng Li, Tianhe Ren, Xueyan Zou, Jianwei Yang, Hang Su, Jun Zhu, et~al.
\newblock Llava-plus: Learning to use tools for creating multimodal agents.
\newblock {\em arXiv preprint arXiv:2311.05437}, 2023.

\bibitem{time_liu2024autotimes}
Yong Liu, Guo Qin, Xiangdong Huang, Jianmin Wang, and Mingsheng Long.
\newblock Autotimes: Autoregressive time series forecasters via large language models.
\newblock {\em arXiv preprint arXiv:2402.02370}, 2024.

\bibitem{liu2021jaqen}
Zaoxing Liu, Hun Namkung, Georgios Nikolaidis, Jeongkeun Lee, Changhoon Kim, Xin Jin, Vladimir Braverman, Minlan Yu, and Vyas Sekar.
\newblock Jaqen: A $\{$High-Performance$\}$$\{$Switch-Native$\}$ approach for detecting and mitigating volumetric $\{$DDoS$\}$ attacks with programmable switches.
\newblock In {\em 30th USENIX Security Symposium (USENIX Security 21)}, pages 3829--3846, 2021.

\bibitem{video_maaz2023video}
Muhammad Maaz, Hanoona Rasheed, Salman Khan, and Fahad~Shahbaz Khan.
\newblock Video-chatgpt: Towards detailed video understanding via large vision and language models.
\newblock {\em arXiv preprint arXiv:2306.05424}, 2023.

\bibitem{pretrain_qiu2020pre}
Xipeng Qiu, Tianxiang Sun, Yige Xu, Yunfan Shao, Ning Dai, and Xuanjing Huang.
\newblock Pre-trained models for natural language processing: A survey.
\newblock {\em Science China Technological Sciences}, 63(10):1872--1897, 2020.

\bibitem{carpet_radware}
Itay Raviv.
\newblock Ddos carpet-bombing – coming in fast and brutal, 2023.
\newblock \url{https://www.radware.com/blog/ddos-protection/2023/07/ddos-carpet-bombing-coming-in-fast-and-brutal}.

\bibitem{promptengineering_sahoo2024systematic}
Pranab Sahoo, Ayush~Kumar Singh, Sriparna Saha, Vinija Jain, Samrat Mondal, and Aman Chadha.
\newblock A systematic survey of prompt engineering in large language models: Techniques and applications.
\newblock {\em arXiv preprint arXiv:2402.07927}, 2024.

\bibitem{pos_schmid1994part}
Helmut Schmid.
\newblock Part-of-speech tagging with neural networks.
\newblock {\em arXiv preprint cmp-lg/9410018}, 1994.

\bibitem{cic_sharafaldin2019developing}
Iman Sharafaldin, Arash~Habibi Lashkari, Saqib Hakak, and Ali~A Ghorbani.
\newblock Developing realistic distributed denial of service (ddos) attack dataset and taxonomy.
\newblock In {\em 2019 international carnahan conference on security technology (ICCST)}, pages 1--8. IEEE, 2019.

\bibitem{touvron2023llama}
Hugo Touvron, Louis Martin, Kevin Stone, Peter Albert, Amjad Almahairi, Yasmine Babaei, Nikolay Bashlykov, Soumya Batra, Prajjwal Bhargava, Shruti Bhosale, et~al.
\newblock Llama 2: Open foundation and fine-tuned chat models.
\newblock {\em arXiv preprint arXiv:2307.09288}, 2023.

\bibitem{transformer}
Ashish Vaswani, Noam Shazeer, Niki Parmar, Jakob Uszkoreit, Llion Jones, Aidan~N Gomez, {\L}ukasz Kaiser, and Illia Polosukhin.
\newblock Attention is all you need.
\newblock {\em Advances in neural information processing systems}, 30, 2017.

\bibitem{wichtlhuber2022ixp}
Matthias Wichtlhuber, Eric Strehle, Daniel Kopp, Lars Prepens, Stefan Stegmueller, Alina Rubina, Christoph Dietzel, and Oliver Hohlfeld.
\newblock Ixp scrubber: learning from blackholing traffic for ml-driven ddos detection at scale.
\newblock In {\em Proceedings of the ACM SIGCOMM 2022 Conference}, pages 707--722, 2022.

\bibitem{network_wu2024large}
Duo Wu, Xianda Wang, Yaqi Qiao, Zhi Wang, Junchen Jiang, Shuguang Cui, and Fangxin Wang.
\newblock Large language model adaptation for networking.
\newblock {\em arXiv preprint arXiv:2402.02338}, 2024.

\bibitem{link_xing2021ripple}
Jiarong Xing, Wenqing Wu, and Ang Chen.
\newblock Ripple: A programmable, decentralized $\{$Link-Flooding$\}$ defense against adaptive adversaries.
\newblock In {\em 30th USENIX Security Symposium (USENIX Security 21)}, pages 3865--3881, 2021.

\bibitem{llmsecurity_yan2023prompt}
Pei Yan, Shunquan Tan, Miaohui Wang, and Jiwu Huang.
\newblock Prompt engineering-assisted malware dynamic analysis using gpt-4.
\newblock {\em arXiv preprint arXiv:2312.08317}, 2023.

\bibitem{carpet_a10}
Terry Young.
\newblock Carpet-bombing attacks highlight the need for intelligent and automated ddos protection, 2024.
\newblock \url{https://www.a10networks.com/blog/carpet-bombing-attacks-highlight-the-need-for-intelligent-and-automated-ddos-protection}.

\bibitem{corr_yu2011discriminating}
Shui Yu, Wanlei Zhou, Weijia Jia, Song Guo, Yong Xiang, and Feilong Tang.
\newblock Discriminating ddos attacks from flash crowds using flow correlation coefficient.
\newblock {\em IEEE transactions on parallel and distributed systems}, 23(6):1073--1080, 2011.

\bibitem{zeng2022glm}
Aohan Zeng, Xiao Liu, Zhengxiao Du, Zihan Wang, Hanyu Lai, Ming Ding, Zhuoyi Yang, Yifan Xu, Wendi Zheng, Xiao Xia, et~al.
\newblock Glm-130b: An open bilingual pre-trained model.
\newblock {\em arXiv preprint arXiv:2210.02414}, 2022.

\bibitem{robotic_zeng2023large}
Fanlong Zeng, Wensheng Gan, Yongheng Wang, Ning Liu, and Philip~S Yu.
\newblock Large language models for robotics: A survey.
\newblock {\em arXiv preprint arXiv:2311.07226}, 2023.

\bibitem{survey_zhang2024mm}
Duzhen Zhang, Yahan Yu, Chenxing Li, Jiahua Dong, Dan Su, Chenhui Chu, and Dong Yu.
\newblock Mm-llms: Recent advances in multimodal large language models.
\newblock {\em arXiv preprint arXiv:2401.13601}, 2024.

\bibitem{corr_zhang2012network}
Jun Zhang, Yang Xiang, Yu~Wang, Wanlei Zhou, Yong Xiang, and Yong Guan.
\newblock Network traffic classification using correlation information.
\newblock {\em IEEE Transactions on Parallel and Distributed systems}, 24(1):104--117, 2012.

\bibitem{YaTC_zhao2023yet}
Ruijie Zhao, Mingwei Zhan, Xianwen Deng, Yanhao Wang, Yijun Wang, Guan Gui, and Zhi Xue.
\newblock Yet another traffic classifier: A masked autoencoder based traffic transformer with multi-level flow representation.
\newblock In {\em Proceedings of the AAAI Conference on Artificial Intelligence}, volume~37, pages 5420--5427, 2023.

\bibitem{net-beacon-zhou2023efficient}
Guangmeng Zhou, Zhuotao Liu, Chuanpu Fu, Qi~Li, and Ke~Xu.
\newblock An efficient design of intelligent network data plane.
\newblock In {\em 32nd USENIX Security Symposium (USENIX Security 23)}, pages 6203--6220, 2023.

\bibitem{time_zhou2024one}
Tian Zhou, Peisong Niu, Liang Sun, Rong Jin, et~al.
\newblock One fits all: Power general time series analysis by pretrained lm.
\newblock {\em Advances in neural information processing systems}, 36, 2024.

\bibitem{compression_zhu2023survey}
Xunyu Zhu, Jian Li, Yong Liu, Can Ma, and Weiping Wang.
\newblock A survey on model compression for large language models.
\newblock {\em arXiv preprint arXiv:2308.07633}, 2023.

\bibitem{llmsecurity_ziems2023explaining}
Noah Ziems, Gang Liu, John Flanagan, and Meng Jiang.
\newblock Explaining tree model decisions in natural language for network intrusion detection.
\newblock {\em arXiv preprint arXiv:2310.19658}, 2023.

\end{thebibliography}

\begin{appendices}

\end{appendices}
\end{document}